\documentclass[aps,prb,twocolumn,amsmath,amssymb,superscriptaddress,citeautoscript,longbibliography,floatfix]{revtex4-2}

\usepackage{graphicx}
\usepackage{dcolumn}
\usepackage{bm}
\usepackage{color}
\usepackage{mathptmx}
\usepackage{mathrsfs}
\usepackage{multirow}
\usepackage{natbib}

\usepackage{epsf}
\usepackage{epsfig}
\usepackage{epstopdf}

\newcommand{\MN}[2]{{\color{black}#2}}                

\begin{document}

\title{Effect of chemical and hydrostatic pressure on the coupled magnetostructural transition of Ni-Mn-In Heusler alloys}

\author{P.\ Devi}
\altaffiliation{Present address: Division of Materials Science \& Engineering, Ames Laboratory, Ames, IA 50011, USA}
\email{parul.devi@cpfs.mpg.de}
\affiliation{Max Planck Institute for Chemical Physics of Solids,  N\"{o}thnitzer Str.\ 40, 01187 Dresden, Germany}

\author{C.\ Salazar Mej\'{i}a}
\affiliation{Dresden High Magnetic Field Laboratory (HLD-EMFL), Helmholtz-Zentrum Dresden-Rossendorf, 01328 Dresden, Germany}

\author{L.\ Caron}
\affiliation{Faculty of Physics, Bielefeld University, 33615 Bielefeld, Germany}

\author{Sanjay Singh}
\affiliation{Max Planck Institute for Chemical Physics of Solids,  N\"{o}thnitzer Str.\ 40, 01187 Dresden, Germany}
\affiliation{School of Materials Science and Technology, Indian Institute of Technology (BHU), Varanasi-221005, India}

\author{M.\ Nicklas}
\email{nicklas@cpfs.mpg.de}
\affiliation{Max Planck Institute for Chemical Physics of Solids,  N\"{o}thnitzer Str.\ 40, 01187 Dresden, Germany}

\author{C.\ Felser}
\affiliation{Max Planck Institute for Chemical Physics of Solids,  N\"{o}thnitzer Str.\ 40, 01187 Dresden, Germany}

\date{\today}

\begin{abstract}
Ni-Mn-In magnetic shape-memory Heusler alloys exhibit generally a large thermal hysteresis at their first-order martensitic phase transition which hinder a technological application in magnetic refrigeration. By optimizing the Cu content in Ni$_2$Cu$_x$Mn$_{1.4-x}$In$_{0.6}$, we obtained a thermal hysteresis of the martensitic phase transition in Ni$_{2}$Cu$_{0.2}$Mn$_{1.2}$In$_{0.6}$ of only 6~K. We can explain this very small hysteresis by an almost perfect habit plane at the interface of martensite and austenite phases. Application of hydrostatic pressure does not reduce the hysteresis further, but shifts the martensitic transition close to room temperature. The isothermal entropy change does not depend on warming or cooling protocols and is pressure independent. Experiments in pulsed-magnetic fields on Ni$_{2}$Cu$_{0.2}$Mn$_{1.2}$In$_{0.6}$ find a reversible magnetocaloric effect with a maximum adiabatic temperature change of $-13$~K.

\end{abstract}

\maketitle


In recent past, tremendous efforts have been made to replace the conventional vapor-based refrigeration by magnetic cooling technology based on the magnetocaloric effect (MCE), which has the potential of lower costs and being more environmental-friendly \cite{Franco2012, Guillou2018}. The MCE manifests itself as a change in the temperature of a material  exposed to a change of magnetic field. It is quantified in terms of an isothermal entropy or an adiabatic temperature change \cite{Caron2009, Zavareh2015}. One material class with very promising properties are the magnetic shape-memory Heusler compounds \cite{Devi2018, Nayak2009, Caron2017, Liu2012, Devi2019}. Much effort is currently devoted to reduce the thermal hysteresis in Heusler materials in order to exploit the large magnetic moment change and the sharp martensitic transition near room temperature that these materials exhibit and, in this way, to obtain a large and reversible MCE. This can be done up to a certain extent by applying either chemical substitution (pressure) or external (hydrostatic) pressure. A reduction of the size of the thermal and magnetic hysteresis is then a result of an improved compatibility between austenite and martensite phases by changes in interatomic distances \cite{ Caron2018, Devi2018a, Devi2019, Liu2012, Song2013}.

Intermetallic Heusler compounds are an exciting class of materials due to their multifunctional properties, such as giant magnetocaloric effect \cite{Liu2012, Krenke2005}, large zero-field cooled exchange bias \cite{Wang2011, Nayak2013}, giant tunable exchange bias \cite{Nayak2015}, spin-glass behavior \cite{chatterjee2009}, large magnetoresistance \cite{Singh2012}, large magnetostriction from a modulated structure \cite{Kainuma2006, SalazarMejia2015, Singh2017}, magnetic antiskyrmions \cite{Nayak2017}, and large canting angles between the magnetic moments \cite{Mescheriakova2014}. These properties can be achieved via tuning multiple parameters, such as number of valence electrons, atomic positions, degree of atomic disorder, and the type and strength of exchange interactions between the atoms in the flexible structure of the Heusler compounds \cite{Nayak2014}. In particular, Ni-Mn based magnetic shape memory Heusler compounds are the subject of special interest thanks to the coexistence of structural and magnetic transitions from a high temperature cubic austenite phase to a low temperature martensite phase, which makes these materials promising candidates for applications as magnetic actuators and energy conversion devices \cite{Gueltig2014,Srivastava2011}. In particular, the thermal hysteresis, Curie and martensitic transition temperatures, martensitic structure, field-induced strain, magnetocrystalline anisotropy and other material properties in the Heusler compounds are extremely sensitive to their composition \cite{Singh2017, Singh2015, Gottschall2015, Caron2017, Khovaylo2009, Oikawa2006, Sharma2010}.

Recent studies provide evidence for an influence of chemical and hydrostatic pressure on the size of the thermal hysteresis and reversibility of conventional and inverse MCE in Ni-Mn based magnetic shape-memory Heusler compounds \cite{Devi2018a, Devi2019, Caron2018}. In the Ni-Mn-In magnetic shape-memory Heusler materials, the Ruderman-Kittel-Kasuya-Yosida interaction between neighboring Mn atoms plays an important role for establishing ferromagnetism in the austenite phase. The driving force for maintaining the instability of the cubic austenite phase is the Jahn-Teller splitting which arises due to the hybridization of the Ni $3d$ states and the $3d$ states of antiferromagnetically coupled Mn atoms at In sites. Any variation in the stoichiometry or substitution of another element which has different valence electrons in the same state affects the hybridization between Ni and In, resulting in changed properties of the martensitic phase transition \cite{Ye2010, Sharma2010, Khan2016}.

In the present Rapid Communication, we investigate the effect of both chemical and hydrostatic pressure on the thermal hysteresis, magnetic-entropy change, magnetostriction, and MCE in Ni$_2$Cu$_x$Mn$_{1.4-x}$In$_{0.6}$. By optimizing the Cu content, we reduced the size of the thermal hysteresis down to 6~K in Ni$_{2}$Cu$_{0.2}$Mn$_{1.2}$In$_{0.6}$, which also shows the largest ordered moment in the martensitic phase in the whole series. The Cu substitution causes a large reduction of the thermal hysteresis, which indicates the approach of compatibility of austenite and martensite phases by forming an almost perfect habit plane at their interface\MN{}{, \textit{i.e.} an undistorted interface between the parent austenite and the low temperature martensite phase}. The deviation from the ideal compatibility condition is only 0.02\% for $x=0.2$. To further optimize the properties we applied hydrostatic pressure on Ni$_{2}$Cu$_{0.2}$Mn$_{1.2}$In$_{0.6}$. Contrary to our expectation and previous studies on Ni-Mn-In magnetic shape memory Heusler compounds \cite{Liu2012, Caron2018, Manosa2008}, application of hydrostatic pressure does not reduce thermal hysteresis anymore. However, we observe a strong shift of the martensitic transition up to room temperature.
We can relate the pressure-induced shift of martensitic transition and the unaffected size of the thermal hysteresis to an enhancement of hybridization of $3d^{10}$ states of Cu atoms at the antiferromagnetically coupled Mn atoms at In sites. The entropy change during warming and cooling protocols \MN{}{is} almost independent of the applied pressure. The field-induced magnetocaloric transition in Ni$_{2}$Cu$_{0.2}$Mn$_{1.2}$In$_{0.6}$ displays a large magnetostriction and a reversible MCE.


Polycrystalline ingots of Ni-Cu-Mn-In were prepared by arc-melting off-stoichiometric amounts of the constituent elements under argon atmosphere and subsequently annealed for 3 days followed by quenching in an ice/water mixture. The composition of the prepared ingot was determined by energy dispersive x-ray analysis. The temperature dependent x-ray diffraction (XRD) experiments were conducted on annealed powder to reduce the residual stress generated during grinding using a Huber G670 camera (Guinier technique, $\lambda = 1.54056$\,\AA~Cu-$K_{\alpha1}$ radiation). The magnetic properties were investigated utilizing physical and magnetic property measurement systems (PPMS and MPMS, Quantum Design). Magnetization measurements under hydrostatic pressure were performed in a home-made CuBe piston-cylinder-type pressure cell built to fit in the sample space of the MPMS. A piece of the Ni$_2$Cu$_x$Mn$_{1.4-x}$In$_{0.6}$ ingot together with a small piece of Sn as manometer was loaded in the pressure cell. At low temperatures the superconducting transition of Sn was used to deduce the pressure inside the cell \cite{Eiling1981}.
It is worth noticing that, there is a pressure drop in the pressure cell due to cooling. Therefore, the pressures stated on this work, which were obtained from the superconducting transition of Sn (3.7~K at ambient pressure), are actually slightly lower than the pressures at which the sample was measured around room temperature. However, this pressure difference does not affect the conclusions drawn in this paper since the rate at which the transition temperatures shift with pressure is unaffected by the shift of the absolute values due to the pressure drop.
Pulsed magnetic field experiments were performed at the Dresden High Magnetic Field Laboratory using a home-built set up for the magnetocaloric measurements \cite{Zavareh2015} and a resistive strain-gauge technique for the magnetostriction experiments.

\begin{figure}[t!]
\includegraphics[width=0.9\linewidth]{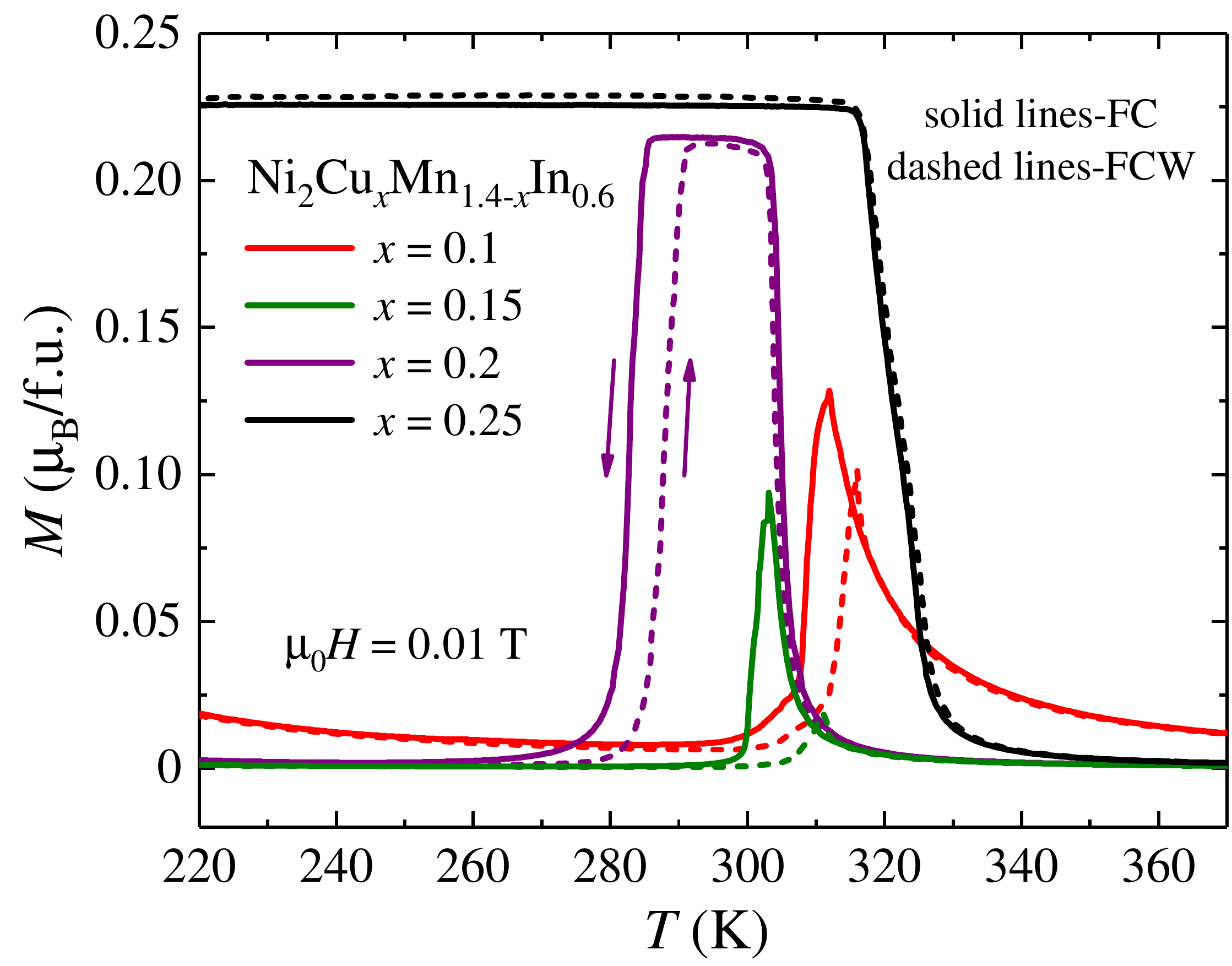}
\centering
\caption{Temperature dependence of the magnetization $M(T)$ at 0.01~T for four different compositions of Ni$_2$Cu$_x$Mn$_{1.4-x}$In$_{0.6}$ measured during FC and FCW.
}
\label{Magnetization}
\end{figure}

Figure \ref{Magnetization} displays the temperature dependence of the magnetization $M(T)$ for four different compositions of Ni$_{2}$Cu$_{x}$Mn$_{1.4-x}$In$_{0.6}$, $x=0.1$, 0.15, 0.2 and 0.25, measured at 0.01~T during cooling in magnetic field, field cooling (FC) protocol, and subsequently upon heating to the starting temperature gain, field-cooled warming (FCW) protocol. Cu substitution up to $x=0.2$ results in a shift of the martensitic transition to lower temperatures. At a little higher Cu substitution level of $x = 0.25$ only the ferromagnetic ordering is maintained without any structural transition resulting in a slightly higher Curie temperature and a larger magnetic moment, similar to the Co$_{2}$-based ferromagnetic Heusler compounds in which $T_C\propto M$ was found \cite{Wollmann2017}. The obtained Curie temperatures $T_C$, the martensitic transition temperatures upon cooling $T_{A-M}$ and warming $T_{M-A}$\MN{}{, and the width of thermal hysteresis $\Delta T_{hyst}$, obtained form the austenitic and martensitic start ($A_{s}$ and $M_{s}$) and finish temperatures ($A_{f}$ and $M_{f}$) by $\Delta T_{hyst}=[(A_s+A_f)-(M_s+M_f)]/{2}$,} are given in Table \ref{tab}. \MN{}{In Ni$_{2}$Cu$_{0.2}$Mn$_{1.2}$In$_{0.6}$ we find a shift of the martensitic transition temperature with magnetic field of about 2.5~K/T.}

\begin{table}[h!]
\begin{center}
\begin{tabular}{ c c c c c }
$x$ &\hspace{0.5 cm}${0.1}$\hspace{0.5 cm}&\hspace{0.5 cm}${0.15}$\hspace{0.5 cm}&\hspace{0.5 cm}${0.2}$\hspace{0.5 cm}&\hspace{0.5 cm}${0.25}$\hspace{0.5 cm}\\\hline
$T_{M-A}$ (K)&\hspace{0.5 cm}312\hspace{0.5 cm}&\hspace{0.5 cm}307\hspace{0.5 cm}&\hspace{0.5 cm}284\hspace{0.5 cm}&\hspace{0.5 cm}$-$\hspace{0.5 cm}\\
$T_{A-M}$ (K)&\hspace{0.5 cm}305\hspace{0.5 cm}&\hspace{0.5 cm}303\hspace{0.5 cm}&\hspace{0.5 cm}282\hspace{0.5 cm}&\hspace{0.5 cm}$-$\hspace{0.5 cm}\\
$T_{C}$ (K)&\hspace{0.5 cm}317\hspace{0.5 cm}&\hspace{0.5 cm}312\hspace{0.5 cm}&\hspace{0.5 cm}303\hspace{0.5 cm}&\hspace{0.5 cm}318\hspace{0.5 cm}\\
\MN{}{$\Delta T_{hyst}$ (K)}&\hspace{0.5 cm}\MN{}{8}\hspace{0.5 cm}&\hspace{0.5 cm}\MN{}{7}\hspace{0.5 cm}&\hspace{0.5 cm}\MN{}{6}\hspace{0.5 cm}&\hspace{0.5 cm}$-$\hspace{0.5 cm}\\
\end{tabular}
\end{center}
\caption{Characteristic temperatures of the investigated Ni$_2$Cu$_x$Mn$_{1.4-x}$In$_{0.6}$ samples. $T_{M-A}$ represents the transition temperature from the martensite to the austenite upon warming and $T_{A-M}$ the corresponding transition temperature upon cooling. $T_{C}$ is the Curie temperature in the austenite phase \MN{}{and $\Delta T_{hyst}$ the width of the thermal hysteresis.}}
\label{tab}
\end{table}

\begin{figure}[t!]
\includegraphics[width=0.9\linewidth]{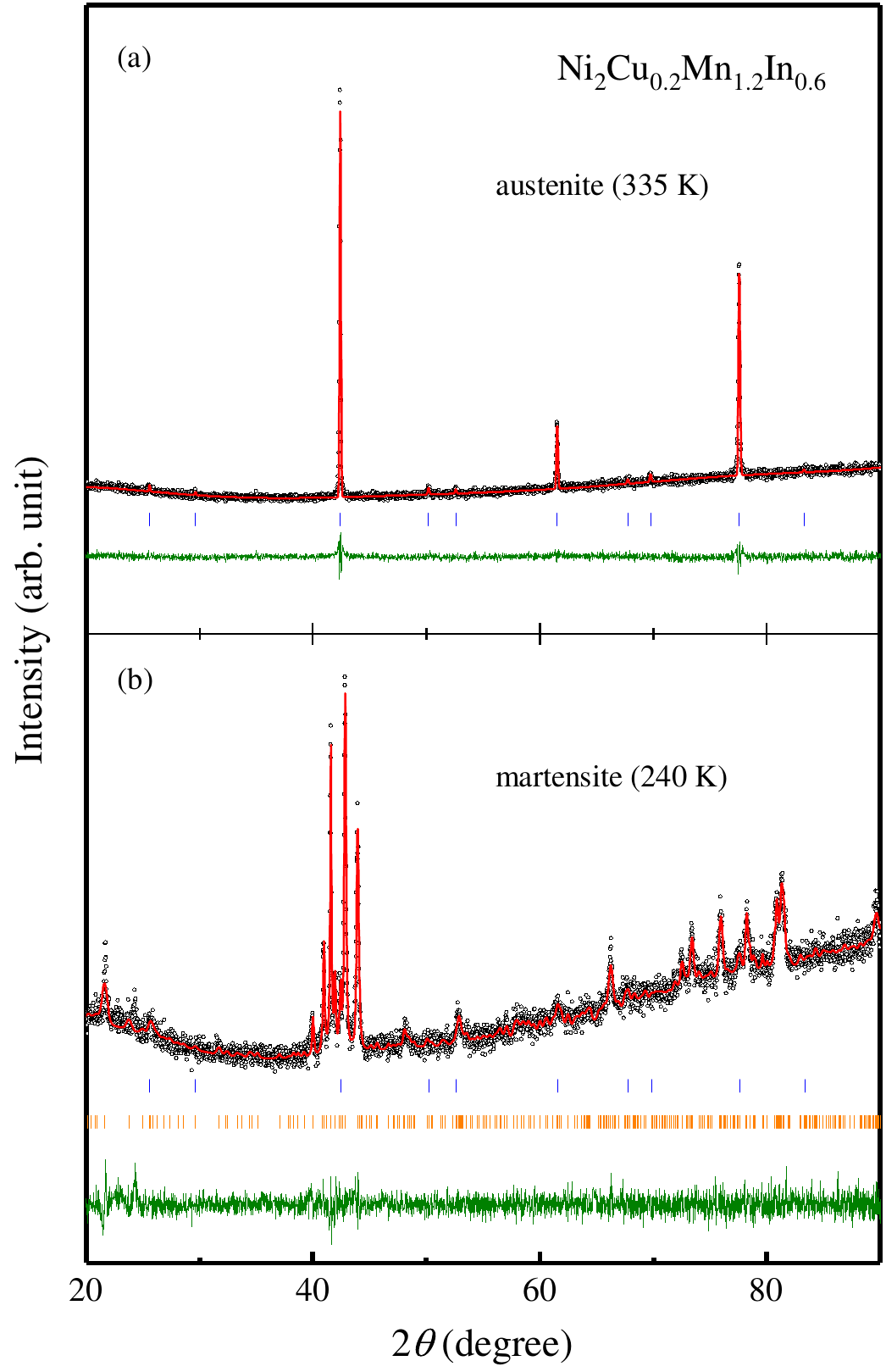}
\centering
\caption{Diffraction patterns of Ni$_{2}$Cu$_{0.2}$Mn$_{1.2}$In$_{0.6}$ collected in the (a) austenite (335~K) and the (b) martensite (240~K) phase. The experimental data, fitted curves and residues are shown by black circles and red and green lines, respectively. Blue ticks mark the Bragg-peak positions corresponding to the L2$_{1}$ cubic structure and the orange ticks the ones corresponding to the $3M$ modulated monoclinic structure.}
\label{XRD}
\end{figure}

The observation of a sharp martensitic transition with a very small thermal hysteresis of about 6~K and  high magnetic moment favors a large MCE. \MN{}{We note that our value for the size of the hysteresis is comparable with the best values found in other Heusler compounds, such as Ni$_{46}$Co$_{3}$Mn$_{35}$Cu$_{2}$In$_{14}$ \cite{Li2019}, Ni$_{43.5}$Co$_{6.5}$Mn$_{39.5}$Sn$_{10.5}$ \cite{Qu2019}, Ni$_{50.7}$Mn$_{33.4}$In$_{15.6}$V$_{0.3}$ \cite{Liu19}, Ni$_{51}$Mn$_{33.4}$In$_{15.6}$ \cite{Stern14}.} Therefore, we studied the geometric compatibility of the austenite and martensite structures in Ni$_{2}$Cu$_{0.2}$Mn$_{1.2}$In$_{0.6}$. XRD patterns were recorded in both phases as shown in Fig.\ \ref{XRD}. The austenite phase exhibits an L2$_{1}$ cubic structure (space group $Fm{\rm-}3m$) with lattice constant $a = 6.0231$~\AA, while the martensite phase shows a $3M$ modulated monoclinic structure (space group $P$2/$m$) with lattice constants $a = 4.4104$~\AA, $b = 5.6423$~\AA, $c = 13.0449$~\AA, and $\beta = 93.0208^{\circ}$. We note that a small fraction of the cubic phase coexists at 240~K due to residual stress which could be generated upon grinding the ingot into powder. Chemical pressure modifies the lattice constants but the modulated structure is similar to that reported in the Ni-Mn-In Heusler family at ambient pressure \cite{Devi2018}. The middle eigenvalue of the transformation matrix between the martensite and the austenite phases, calculated using the lattice parameters above, is 0.9998. That is a deviation of only $0.02\%$ from unity. This is one of the smallest deviation of the middle eigenvalue till date observed in magnetic shape-memory Heusler compounds \cite{Liu2012, Stern14, Caron2018, Devi2019, Liu19}. The closer this value is to unity, the better is the compatibility between austenite and martensite structures. As a result the interface between austenite and martensite phases consists of an almost perfect habit plane, which leads to a better reversibility of the structural transition and a reduced hysteresis.

\begin{figure}[t!]
\includegraphics[width=0.9\linewidth]{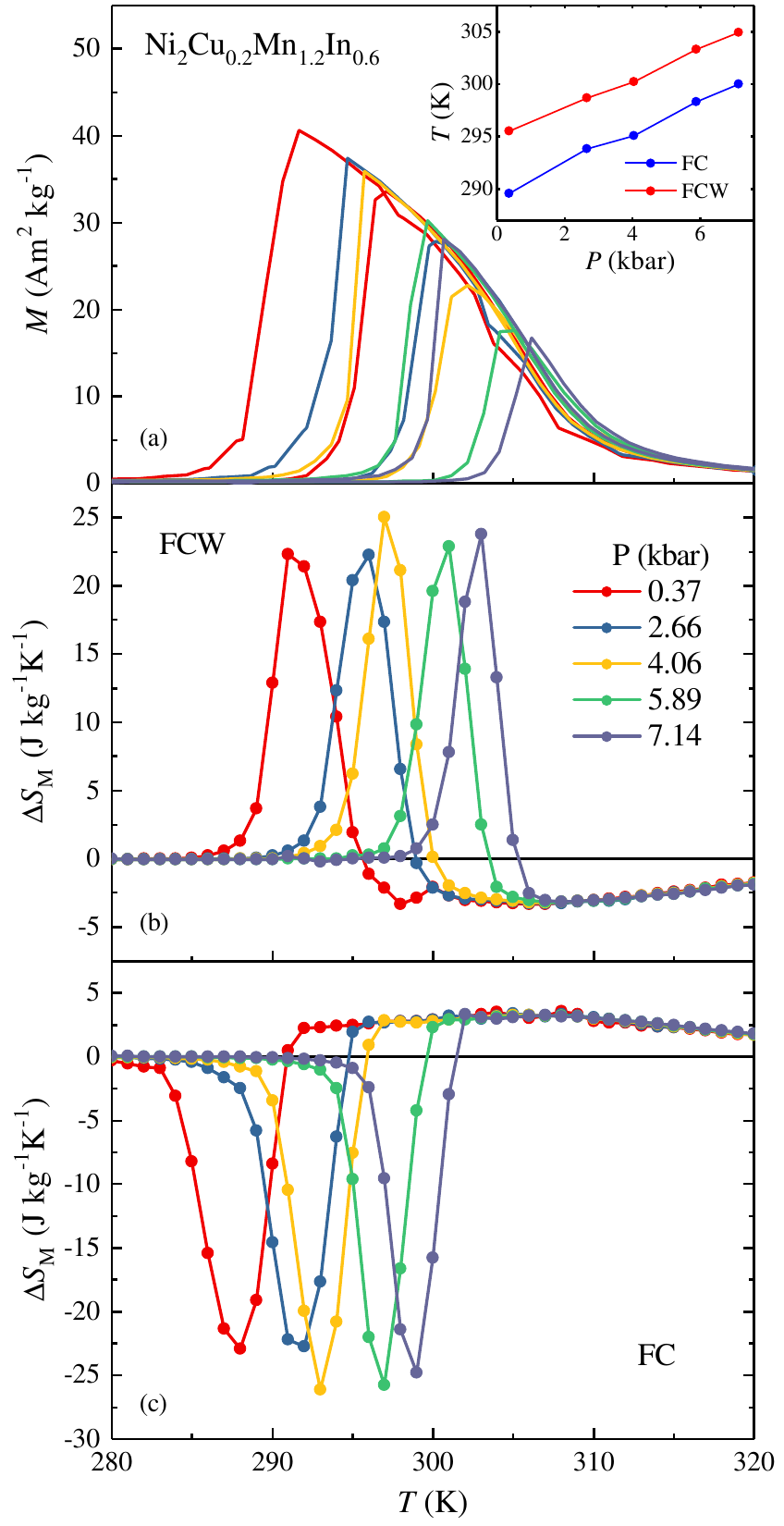}
\centering
\caption{(a) $M(T)$ curves recorded during FC and FCW protocols under different applied pressures in a magnetic field of 0.01~T for Ni$_{2}$Cu$_{0.2}$Mn$_{1.2}$In$_{0.6}$. The inset shows the shift of the martensitic phase transition with pressure. (b) and (c) Magnetic entropy change for a magnetic field change from 0 to 2~T at different applied pressures for warming and cooling protocols, respectively.}
\label{Pressure magnetization}
\end{figure}

A reduction in the size of the hysteresis upon application of external pressure has been observed in a number of compounds belonging to the family of the Ni-Mn-based magnetic shape-memory alloys, such as Ni$_{45.7}$Co$_{4.2}$Mn$_{36.6}$In$_{13.5}$ \cite{Gottschall2016a}, Ni$_{45.2}$Co$_{5.1}$Mn$_{36.7}$In$_{13}$ \cite{Liu2012}, Ni$_{50}$Mn$_{35}$In$_{15}$ \cite{Manosa2008}, Ni$_{45}$Co$_{5}$Mn$_{38}$Sb$_{12}$ \cite{Nayak2009}. This suggested the possibility of a further reduction of the already small hysteresis of the martensitic transformation in Ni$_{2}$Cu$_{0.2}$Mn$_{1.2}$In$_{0.6}$ by external pressure. In contrast to that, we found that application of pressure shifts the martensitic transition in Ni$_{2}$Cu$_{0.2}$Mn$_{1.2}$In$_{0.6}$ toward higher temperature, but without affecting the size of the thermal hysteresis significantly (see Fig.\ \ref{Pressure magnetization}(a)).
The martensitic transition temperature depends almost linearly on pressure with slopes of $dT_{M-A}/dp \approx 1.86$~K/kbar and $dT_{A-M}/dp \approx 1.98$~K/kbar upon warming and cooling, respectively (see inset of Fig.\ \ref{Pressure magnetization}a). Simultaneously, the Curie temperature remains almost unchanged.
As a consequence the maximum change in magnetization decreases upon increasing pressure, since the ferromagnetic ordering in the austenitic phase is not completed before the transition to the martensitic phase takes place upon decreasing temperature.
Since the martensitic phase possesses a smaller volume than the austenitic phase, applied hydrostatic pressure stabilizes the martensitic phase. That explains the shift of the martensitic transition toward higher temperatures by an enhancement of the hybridization between the 3$d$ and 3$p$ states of Mn and Cu \cite{Ye2010, Khan2016}.
The results indicate that increasing pressure does not affect the formation of the habit plane between austenite and martensite phases in Ni$_{2}$Cu$_{0.2}$Mn$_{1.2}$In$_{0.6}$.

We recorded the temperature dependence of the magnetization $M(T)$ at different applied fields up to 2~T in order to calculated the isothermal entropy change $\Delta S_M$ using the Maxwell relation: $\Delta S_{M} = S (T, H) - S (T, 0)= \int_{0}^{H} \left({\partial M (T, H)\over\partial T}\right)_H dH$. At ambient pressure we find a maximum isothermal entropy change $\Delta S_M\approx 22.5$~J\,kg$^{-1}$K$^{-1}$ upon warming and of $-25$~J\,kg$^{-1}$K$^{-1}$ upon cooling. Figures \ref{Pressure magnetization}(b) and \ref{Pressure magnetization}(c) display the $\Delta S_M(T)$ upon warming and cooling for different applied pressures. Within the error bars, $|\Delta S_M(T)|$ exhibits similar maximum values for data recorded for different applied pressures upon warming and cooling. \MN{}{The maximum difference between highest and lowest value of $|\Delta S_M(T)|$ at different pressures is about 2~J\,kg$^{-1}$K$^{-1}$}.

\begin{figure}[t!]
\includegraphics[width=0.9\linewidth]{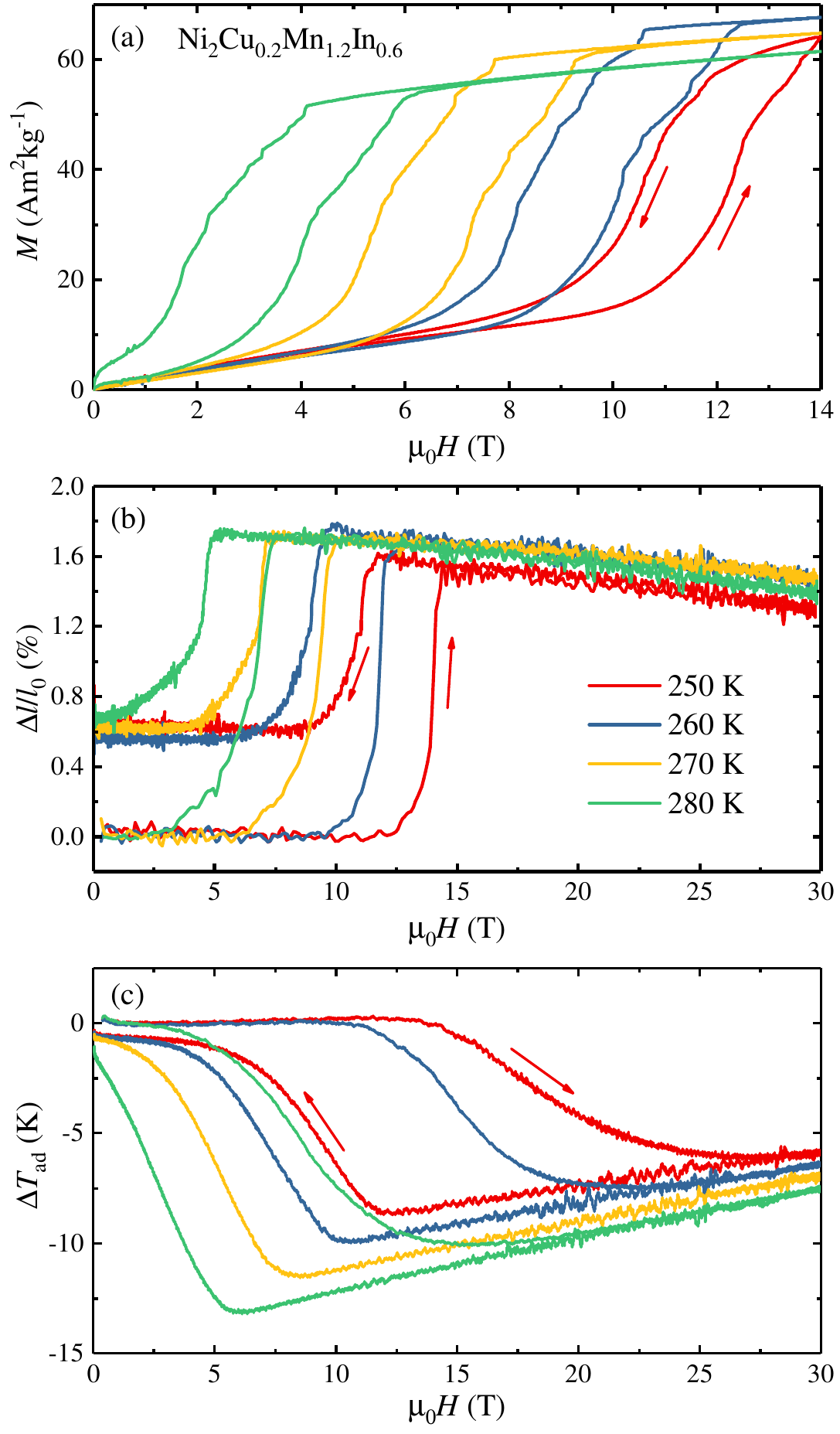}
\centering
\caption{(a) Isothermal magnetization data $M(H)$ measured in static magnetic fields and (b) relative length change $\Delta l/l_0(H)$ and (c) adiabatic temperature change both recorded in pulsed magnetic fields. All data are shown for the same temperatures below $T_{M-A}=284$~K. Before each measurement the sample was warmed up to the austenitic state followed by cooling down to the martensitic state before the target temperature was approached upon warming. The arrows indicate the direction of the field sweeps.}
\label{HighFields}
\end{figure}

So far we have shown that the magnetocaloric properties of Ni$_2$Cu$_x$Mn$_{1.4-x}$In$_{0.6}$ can be optimized by substituting Mn by Cu. In that way we obtained the lowest thermal hysteresis, an almost ideal compatibility condition, and the reversibility of the isothermal entropy change for $x=0.2$. Furthermore, application of hydrostatic pressure on Ni$_{2}$Cu$_{0.2}$Mn$_{1.2}$In$_{0.6}$ shifts its martensitic transition to room temperature without changing the size of the hysteresis and the magnitude of the isothermal entropy change. Motivated by these promising results, we now turn to detailed studies of the isothermal magnetization in static fields and of the magnetostriction and adiabatic temperature change in pulsed magnetic fields on Ni$_{2}$Cu$_{0.2}$Mn$_{1.2}$In$_{0.6}$. The results are presented in Fig.\ \ref{HighFields}. All experiments were carried out at the same temperatures below the reverse martensitic transition temperature $T_{M-A}$. Before each measurement, the sample first was warmed up to the austenitic phase and then cooled down to the fully martensitic phase followed by warming to the target temperature.


Isothermal magnetization $M(H)$ measurements were carried out in static magnetic fields up to 14~T at temperatures below the reverse martensitic transition temperature $T_{M-A}$ in order to investigate the field-induced reverse martensitic transition from the martensitic to the austenitic phase and to determine the corresponding critical fields.  Since the temperatures were reached upon warming, the sample was always in the fully martensitic state and 14~T were sufficient to observe the full reverse martensitic transition at all temperatures.

The magnetostriction was recorded in magnetic field pulses of 30~T. The relative length change is determined as $\Delta l/l_{0} = (l-l_{0})/l_{0}$, where $l_{0}$ is the length of the sample before the magnetic-field pulse. As expected from the $M(H)$ data, a full reverse martensitic transition is present at the investigated temperatures below $T_{M-A}$. The recorded relative change in length of $\Delta l\approx 1.5$\% at the magnetic field-induced transition is large in comparison with other Heusler compounds \cite{Kainuma2006, Salazar2017, Li2010, Ito2007}. In addition, the data also display a large irreversibility, i.e.\ the contraction of the sample upon field removal is only about 67\% of the expansion upon application of the field. We will discuss this observation below.

As in case of the magnetostriction, the direct adiabatic temperature change $\Delta T_{ad}(H)$ was also determined in magnetic-field pulses of 30~T. We find an almost reversible MCE independent of the initial temperature within the error bars of the experiment \cite{Gottschall17, Engelbrecht13, Guillou14, Skokov13}.
Due to the large magnetic fields, $\Delta T_{ad}(H)$ displays a complex behavior. During the up-sweep of the magnetic field the sample first cools down followed by a weak increase in $\Delta T_{ad}$. A similar behavior is found during the down-sweep, but with a pronounced minimum in $\Delta T_{ad}(H)$ which shifts to lower fields upon increasing the initial temperature of the experiment. Simultaneously, the maximum cooling effect increases upon approaching $T_{M-A}$ to about 13~K at 280~K. In magnetic shape-memory Heusler alloys the MCE due to the reverse martensitic transformation has two main contributions, a structural and a magnetic \cite{Liu2012, Zavareh2015,Gottschall2016a}. The structural contribution leads to an inverse MCE (cooling upon increasing field) and the magnetic contribution to a conventional MCE (warming upon increasing field). The competition of both effects lead to the observed behavior of $\Delta T_{ad}(H)$.

The MCE is reversible at all investigated temperatures below the reverse martensitic transition temperature $T_{M-A}$, in contrast to the observed irreversibility in the magnetostriction data. The reversible MCE indicates that upon removal of the field the sample transforms back to a fully martensitic phase and that no fraction of the austenitic phase is arrested as observed in other Heusler alloys \cite{Nayak2014}. The irreversibility in $\Delta l/l_{0}$ originates, therefore, most likely from an alignment of the martensitic variants along the magnetic field. After reaching the initial temperature following the pre-cooling protocol described above the martensitic variants are randomly oriented. However, on the down-sweep of the field from the field-induced austenitic phase the forming martensitic variants align along the field leading to a final length of the sample significantly larger than the initial length. The contraction of $\Delta l/l_{0}$ from the austenite to field-aligned martensite upon decreasing field is only about $2/3$ of the expansion from randomly-oriented martensite to austenite upon increasing field.


To summarize, we have shown that Cu substitution on the Mn site reduces the thermal hysteresis in the Ni-Mn-In magnetic shape-memory Heusler family. The smallest thermal hysteresis of 6~K was obtained in \MN{}{Ni$_{2}$Cu$_{0.2}$Mn$_{1.2}$In$_{0.6}$.} The strongly reduced hysteresis can be explained solely based on the almost perfect compatibility of austenite and martensite phases as evidenced by our XRD data. While Cu substitution has a strong effect on both martensitic transition temperature and thermal hysteresis, application of hydrostatic pressure only shifts the martensitic transition up in temperature without affecting the formation of the habit plane at the interface of austenite and martensite phases.
At only 7~kbar we find the martensitic transition close to room temperature in \MN{}{Ni$_{2}$Cu$_{0.2}$Mn$_{1.2}$In$_{0.6}$}.
The shift of martensitic transition with pressure is directly related to the smaller unit-cell volume of the martensitic phase as compared to that of the austenitic phase. However, the compatibility condition of cubic austenite and monoclinic martensite structures is not affected by hydrostatic pressure as indicated by the similar values of the maximum isothermal entropy change during warming and cooling protocols at different pressures.

The hysteresis of the martensitic transition is related to the energy barrier between austenitic and martensitic phases. The geometrical compatibility between the phases is just one of many factors which influence the energy barrier and, therefore, the thermal hysteresis. By adjusting the Cu content we already obtained an almost perfect compatibility condition. Application of external pressure can only affect the lattice parameters and in that way adjust compatibility of the martensitic and austenitic phases. Since the compatibility condition is almost perfectly met at ambient pressure, the hysteresis could not be improved further by application of external pressure. However, we still observe a quite pronounced hysteresis which strongly points at other factors contributing to the hysteretic behavior.

Finally, we can conclude from our direct adiabatic measurements of the temperature change \MN{}{across the martensitic transition} that the MCE in \MN{}{ Ni$_{2}$Cu$_{0.2}$Mn$_{1.2}$In$_{0.6}$} is reversible, even though, the magnetostriction is not reversible due to an aligning of martensitic variants in applied magnetic field. Our results help to unveil the nature of the first-order martensitic phase transition in Ni-Mn-In based magnetic shape-memory Heusler alloys and highlight that the mechanism responsible for the reduction of the hysteresis of the martensitic phase transition is not universal and must be investigated thoroughly for different alloy compositions.

\begin{acknowledgments}
This work was financially supported by the ERC Advanced Grant ‘TOPMAT” (No. 742068).  We acknowledge the support of the HLD at HZDR, member of the European Magnetic Field Laboratory (EMFL). S.S.\ thanks Science and Engineering Research Board of India for financial support through Early Career Research Award (Grant No.\ ECR/2017/003186) and the award of Ramanujan Fellowship (Grant No.\ SB/S2/RJN-015/2017). We thank Horst Borrmann for carrying out the XRD measurements.
\end{acknowledgments}

\bibliography{MCE_NiCuMnIn}

\begin{thebibliography}{48}%
\makeatletter
\providecommand \@ifxundefined [1]{%
 \@ifx{#1\undefined}
}%
\providecommand \@ifnum [1]{%
 \ifnum #1\expandafter \@firstoftwo
 \else \expandafter \@secondoftwo
 \fi
}%
\providecommand \@ifx [1]{%
 \ifx #1\expandafter \@firstoftwo
 \else \expandafter \@secondoftwo
 \fi
}%
\providecommand \natexlab [1]{#1}%
\providecommand \enquote  [1]{``#1''}%
\providecommand \bibnamefont  [1]{#1}%
\providecommand \bibfnamefont [1]{#1}%
\providecommand \citenamefont [1]{#1}%
\providecommand \href@noop [0]{\@secondoftwo}%
\providecommand \href [0]{\begingroup \@sanitize@url \@href}%
\providecommand \@href[1]{\@@startlink{#1}\@@href}%
\providecommand \@@href[1]{\endgroup#1\@@endlink}%
\providecommand \@sanitize@url [0]{\catcode `\\12\catcode `\$12\catcode
  `\&12\catcode `\#12\catcode `\^12\catcode `\_12\catcode `\%12\relax}%
\providecommand \@@startlink[1]{}%
\providecommand \@@endlink[0]{}%
\providecommand \url  [0]{\begingroup\@sanitize@url \@url }%
\providecommand \@url [1]{\endgroup\@href {#1}{\urlprefix }}%
\providecommand \urlprefix  [0]{URL }%
\providecommand \Eprint [0]{\href }%
\providecommand \doibase [0]{https://doi.org/}%
\providecommand \selectlanguage [0]{\@gobble}%
\providecommand \bibinfo  [0]{\@secondoftwo}%
\providecommand \bibfield  [0]{\@secondoftwo}%
\providecommand \translation [1]{[#1]}%
\providecommand \BibitemOpen [0]{}%
\providecommand \bibitemStop [0]{}%
\providecommand \bibitemNoStop [0]{.\EOS\space}%
\providecommand \EOS [0]{\spacefactor3000\relax}%
\providecommand \BibitemShut  [1]{\csname bibitem#1\endcsname}%
\let\auto@bib@innerbib\@empty
\bibitem [{\citenamefont {Franco}\ \emph {et~al.}(2012)\citenamefont {Franco},
  \citenamefont {Bl\'{a}zquez}, \citenamefont {Ingale},\ and\ \citenamefont
  {Conde}}]{Franco2012}%
  \BibitemOpen
  \bibfield  {author} {\bibinfo {author} {\bibfnamefont {V.}~\bibnamefont
  {Franco}}, \bibinfo {author} {\bibfnamefont {J.~S.}\ \bibnamefont
  {Bl\'{a}zquez}}, \bibinfo {author} {\bibfnamefont {B.}~\bibnamefont
  {Ingale}},\ and\ \bibinfo {author} {\bibfnamefont {A.}~\bibnamefont
  {Conde}},\ }\bibfield  {title} {\bibinfo {title} {The magnetocaloric effect
  and magnetic refrigeration near room temperature: Materials and models},\
  }\href@noop {} {\bibfield  {journal} {\bibinfo  {journal} {Annu. Rev. Mater.
  Res.}\ }\textbf {\bibinfo {volume} {42}},\ \bibinfo {pages} {305} (\bibinfo
  {year} {2012})}\BibitemShut {NoStop}%
\bibitem [{\citenamefont {Guillou}\ \emph {et~al.}(2018)\citenamefont
  {Guillou}, \citenamefont {Pathak}, \citenamefont {Paudyal}, \citenamefont
  {Mudryk}, \citenamefont {Wilhelm}, \citenamefont {Rogalev},\ and\
  \citenamefont {Pecharsky}}]{Guillou2018}%
  \BibitemOpen
  \bibfield  {author} {\bibinfo {author} {\bibfnamefont {F.}~\bibnamefont
  {Guillou}}, \bibinfo {author} {\bibfnamefont {A.~K.}\ \bibnamefont {Pathak}},
  \bibinfo {author} {\bibfnamefont {D.}~\bibnamefont {Paudyal}}, \bibinfo
  {author} {\bibfnamefont {Y.}~\bibnamefont {Mudryk}}, \bibinfo {author}
  {\bibfnamefont {F.}~\bibnamefont {Wilhelm}}, \bibinfo {author} {\bibfnamefont
  {A.}~\bibnamefont {Rogalev}},\ and\ \bibinfo {author} {\bibfnamefont {V.~K.}\
  \bibnamefont {Pecharsky}},\ }\bibfield  {title} {\bibinfo {title}
  {Non-hysteretic first-order phase transition with large latent heat and giant
  low-field magnetocaloric effect},\ }\href@noop {} {\bibfield  {journal}
  {\bibinfo  {journal} {Nat. Commun.}\ }\textbf {\bibinfo {volume} {9}},\
  \bibinfo {pages} {2925} (\bibinfo {year} {2018})}\BibitemShut {NoStop}%
\bibitem [{\citenamefont {Caron}\ \emph {et~al.}(2009)\citenamefont {Caron},
  \citenamefont {Ou}, \citenamefont {Nguyen}, \citenamefont {Thanh},
  \citenamefont {Tegus},\ and\ \citenamefont {Br\"{u}ck}}]{Caron2009}%
  \BibitemOpen
  \bibfield  {author} {\bibinfo {author} {\bibfnamefont {L.}~\bibnamefont
  {Caron}}, \bibinfo {author} {\bibfnamefont {Z.~Q.}\ \bibnamefont {Ou}},
  \bibinfo {author} {\bibfnamefont {T.~T.}\ \bibnamefont {Nguyen}}, \bibinfo
  {author} {\bibfnamefont {D.~T.~C.}\ \bibnamefont {Thanh}}, \bibinfo {author}
  {\bibfnamefont {O.}~\bibnamefont {Tegus}},\ and\ \bibinfo {author}
  {\bibfnamefont {E.}~\bibnamefont {Br\"{u}ck}},\ }\bibfield  {title} {\bibinfo
  {title} {On the determination of the magnetic entropy change in materials
  with first-order transitions},\ }\href@noop {} {\bibfield  {journal}
  {\bibinfo  {journal} {J. Magn. and Magn. Mater.}\ }\textbf {\bibinfo {volume}
  {321}},\ \bibinfo {pages} {3559} (\bibinfo {year} {2009})}\BibitemShut
  {NoStop}%
\bibitem [{\citenamefont {Ghorbani~Zavareh}\ \emph {et~al.}(2015)\citenamefont
  {Ghorbani~Zavareh}, \citenamefont {Salazar~Mejia}, \citenamefont {Nayak},
  \citenamefont {Skourski}, \citenamefont {Wosnitza}, \citenamefont {Felser},\
  and\ \citenamefont {Nicklas}}]{Zavareh2015}%
  \BibitemOpen
  \bibfield  {author} {\bibinfo {author} {\bibfnamefont {M.}~\bibnamefont
  {Ghorbani~Zavareh}}, \bibinfo {author} {\bibfnamefont {C.}~\bibnamefont
  {Salazar~Mejia}}, \bibinfo {author} {\bibfnamefont {A.~K.}\ \bibnamefont
  {Nayak}}, \bibinfo {author} {\bibfnamefont {Y.}~\bibnamefont {Skourski}},
  \bibinfo {author} {\bibfnamefont {J.}~\bibnamefont {Wosnitza}}, \bibinfo
  {author} {\bibfnamefont {C.}~\bibnamefont {Felser}},\ and\ \bibinfo {author}
  {\bibfnamefont {M.}~\bibnamefont {Nicklas}},\ }\bibfield  {title} {\bibinfo
  {title} {Direct measurements of the magnetocaloric effect in pulsed magnetic
  fields: The example of the {H}eusler alloy
  \text{Ni$_{50}$Mn$_{35}$In$_{15}$}},\ }\href@noop {} {\bibfield  {journal}
  {\bibinfo  {journal} {Appl. Phys. Lett.}\ }\textbf {\bibinfo {volume}
  {106}},\ \bibinfo {pages} {071904} (\bibinfo {year} {2015})}\BibitemShut
  {NoStop}%
\bibitem [{\citenamefont {Devi}\ \emph
  {et~al.}(2018{\natexlab{a}})\citenamefont {Devi}, \citenamefont {Singh},
  \citenamefont {Dutta}, \citenamefont {Manna}, \citenamefont {D$'$Souza},
  \citenamefont {Ikeda}, \citenamefont {Suard}, \citenamefont {Petricek},
  \citenamefont {Simon}, \citenamefont {Werner}, \citenamefont {Chadhov},
  \citenamefont {Parkin}, \citenamefont {Felser},\ and\ \citenamefont
  {Pandey}}]{Devi2018}%
  \BibitemOpen
  \bibfield  {author} {\bibinfo {author} {\bibfnamefont {P.}~\bibnamefont
  {Devi}}, \bibinfo {author} {\bibfnamefont {S.}~\bibnamefont {Singh}},
  \bibinfo {author} {\bibfnamefont {B.}~\bibnamefont {Dutta}}, \bibinfo
  {author} {\bibfnamefont {K.}~\bibnamefont {Manna}}, \bibinfo {author}
  {\bibfnamefont {S.~W.}\ \bibnamefont {D$'$Souza}}, \bibinfo {author}
  {\bibfnamefont {Y.}~\bibnamefont {Ikeda}}, \bibinfo {author} {\bibfnamefont
  {E.}~\bibnamefont {Suard}}, \bibinfo {author} {\bibfnamefont
  {V.}~\bibnamefont {Petricek}}, \bibinfo {author} {\bibfnamefont
  {P.}~\bibnamefont {Simon}}, \bibinfo {author} {\bibfnamefont
  {P.}~\bibnamefont {Werner}}, \bibinfo {author} {\bibfnamefont
  {S.}~\bibnamefont {Chadhov}}, \bibinfo {author} {\bibfnamefont {S.~S.~P.}\
  \bibnamefont {Parkin}}, \bibinfo {author} {\bibfnamefont {C.}~\bibnamefont
  {Felser}},\ and\ \bibinfo {author} {\bibfnamefont {D.}~\bibnamefont
  {Pandey}},\ }\bibfield  {title} {\bibinfo {title} {Adaptive modulation in the
  \text{Ni$_{2}$Mn$_{1.4}$In$_{0.6}$} magnetic shape-memory {H}eusler alloy},\
  }\href@noop {} {\bibfield  {journal} {\bibinfo  {journal} {Phys. Rev. B.}\
  }\textbf {\bibinfo {volume} {97}},\ \bibinfo {pages} {224102} (\bibinfo
  {year} {2018}{\natexlab{a}})}\BibitemShut {NoStop}%
\bibitem [{\citenamefont {Nayak}\ \emph {et~al.}(2009)\citenamefont {Nayak},
  \citenamefont {Suresh}, \citenamefont {Nigam}, \citenamefont {Coelho},\ and\
  \citenamefont {Gama}}]{Nayak2009}%
  \BibitemOpen
  \bibfield  {author} {\bibinfo {author} {\bibfnamefont {A.~K.}\ \bibnamefont
  {Nayak}}, \bibinfo {author} {\bibfnamefont {K.~G.}\ \bibnamefont {Suresh}},
  \bibinfo {author} {\bibfnamefont {A.~K.}\ \bibnamefont {Nigam}}, \bibinfo
  {author} {\bibfnamefont {A.~A.}\ \bibnamefont {Coelho}},\ and\ \bibinfo
  {author} {\bibfnamefont {S.}~\bibnamefont {Gama}},\ }\bibfield  {title}
  {\bibinfo {title} {Pressure induced magnetic and magnetocaloric properties in
  \text{NiCoMnSb} {H}eusler alloy},\ }\href@noop {} {\bibfield  {journal}
  {\bibinfo  {journal} {J. Appl. Phys.}\ }\textbf {\bibinfo {volume} {106}},\
  \bibinfo {pages} {053901} (\bibinfo {year} {2009})}\BibitemShut {NoStop}%
\bibitem [{\citenamefont {Caron}\ \emph {et~al.}(2017)\citenamefont {Caron},
  \citenamefont {Dutta}, \citenamefont {Devi}, \citenamefont {Zavareh},
  \citenamefont {Hickel}, \citenamefont {Cabassi}, \citenamefont {Bolzoni},
  \citenamefont {Fabbrici}, \citenamefont {Albertini}, \citenamefont {Felser},\
  and\ \citenamefont {Singh}}]{Caron2017}%
  \BibitemOpen
  \bibfield  {author} {\bibinfo {author} {\bibfnamefont {L.}~\bibnamefont
  {Caron}}, \bibinfo {author} {\bibfnamefont {B.}~\bibnamefont {Dutta}},
  \bibinfo {author} {\bibfnamefont {P.}~\bibnamefont {Devi}}, \bibinfo {author}
  {\bibfnamefont {M.~G.}\ \bibnamefont {Zavareh}}, \bibinfo {author}
  {\bibfnamefont {T.}~\bibnamefont {Hickel}}, \bibinfo {author} {\bibfnamefont
  {R.}~\bibnamefont {Cabassi}}, \bibinfo {author} {\bibfnamefont
  {F.}~\bibnamefont {Bolzoni}}, \bibinfo {author} {\bibfnamefont
  {S.}~\bibnamefont {Fabbrici}}, \bibinfo {author} {\bibfnamefont
  {F.}~\bibnamefont {Albertini}}, \bibinfo {author} {\bibfnamefont
  {C.}~\bibnamefont {Felser}},\ and\ \bibinfo {author} {\bibfnamefont
  {S.}~\bibnamefont {Singh}},\ }\bibfield  {title} {\bibinfo {title} {Effect of
  {P}t substitution on the magnetocrystalline anisotropy of
  \text{Ni$_{2}$MnGa}: A competition between chemistry and elasticity},\
  }\href@noop {} {\bibfield  {journal} {\bibinfo  {journal} {Phys. Rev. B}\
  }\textbf {\bibinfo {volume} {96}},\ \bibinfo {pages} {054105} (\bibinfo
  {year} {2017})}\BibitemShut {NoStop}%
\bibitem [{\citenamefont {Liu}\ \emph {et~al.}(2012)\citenamefont {Liu},
  \citenamefont {Gottschall}, \citenamefont {Skokov}, \citenamefont {Moore},\
  and\ \citenamefont {Gutfleish}}]{Liu2012}%
  \BibitemOpen
  \bibfield  {author} {\bibinfo {author} {\bibfnamefont {J.}~\bibnamefont
  {Liu}}, \bibinfo {author} {\bibfnamefont {T.}~\bibnamefont {Gottschall}},
  \bibinfo {author} {\bibfnamefont {K.~P.}\ \bibnamefont {Skokov}}, \bibinfo
  {author} {\bibfnamefont {J.~D.}\ \bibnamefont {Moore}},\ and\ \bibinfo
  {author} {\bibfnamefont {O.}~\bibnamefont {Gutfleish}},\ }\bibfield  {title}
  {\bibinfo {title} {Giant magnetocaloric effect driven by structural
  transitions},\ }\href@noop {} {\bibfield  {journal} {\bibinfo  {journal}
  {Nat. Mater.}\ }\textbf {\bibinfo {volume} {11}},\ \bibinfo {pages} {620}
  (\bibinfo {year} {2012})}\BibitemShut {NoStop}%
\bibitem [{\citenamefont {Devi}\ \emph {et~al.}(2019)\citenamefont {Devi},
  \citenamefont {Salazar~Mej\'{i}a}, \citenamefont {Zavareh}, \citenamefont
  {Dubey}, \citenamefont {Kushwaha}, \citenamefont {Skourski}, \citenamefont
  {Felser}, \citenamefont {Nicklas},\ and\ \citenamefont {Singh}}]{Devi2019}%
  \BibitemOpen
  \bibfield  {author} {\bibinfo {author} {\bibfnamefont {P.}~\bibnamefont
  {Devi}}, \bibinfo {author} {\bibfnamefont {C.}~\bibnamefont
  {Salazar~Mej\'{i}a}}, \bibinfo {author} {\bibfnamefont {M.~G.}\ \bibnamefont
  {Zavareh}}, \bibinfo {author} {\bibfnamefont {K.~K.}\ \bibnamefont {Dubey}},
  \bibinfo {author} {\bibfnamefont {P.}~\bibnamefont {Kushwaha}}, \bibinfo
  {author} {\bibfnamefont {Y.}~\bibnamefont {Skourski}}, \bibinfo {author}
  {\bibfnamefont {C.}~\bibnamefont {Felser}}, \bibinfo {author} {\bibfnamefont
  {M.}~\bibnamefont {Nicklas}},\ and\ \bibinfo {author} {\bibfnamefont
  {S.}~\bibnamefont {Singh}},\ }\bibfield  {title} {\bibinfo {title} {Improved
  magnetostructural and magnetocaloric reversibility in magnetic
  \text{Ni-Mn-In} shape-memory {H}eusler alloy by optimizing the geometric
  compatibility condition},\ }\href@noop {} {\bibfield  {journal} {\bibinfo
  {journal} {Phys. Rev. Mat.}\ }\textbf {\bibinfo {volume} {3}},\ \bibinfo
  {pages} {062401(R)} (\bibinfo {year} {2019})}\BibitemShut {NoStop}%
\bibitem [{\citenamefont {Caron}\ \emph {et~al.}()\citenamefont {Caron},
  \citenamefont {Devi}, \citenamefont {Carvalho}, \citenamefont {Felser},\ and\
  \citenamefont {Singh}}]{Caron2018}%
  \BibitemOpen
  \bibfield  {author} {\bibinfo {author} {\bibfnamefont {L.}~\bibnamefont
  {Caron}}, \bibinfo {author} {\bibfnamefont {P.}~\bibnamefont {Devi}},
  \bibinfo {author} {\bibfnamefont {A.~M.~G.}\ \bibnamefont {Carvalho}},
  \bibinfo {author} {\bibfnamefont {C.}~\bibnamefont {Felser}},\ and\ \bibinfo
  {author} {\bibfnamefont {S.}~\bibnamefont {Singh}},\ }\bibfield  {title}
  {\bibinfo {title} {Minimizing hysteresis in phase transforming magnetocaloric
  {H}eusler alloys},\ }\href@noop {} {\bibinfo  {journal} {arXiv:1806.05075
  [cond-mat.mtrl-sci]}\ }\BibitemShut {NoStop}%
\bibitem [{\citenamefont {Devi}\ \emph
  {et~al.}(2018{\natexlab{b}})\citenamefont {Devi}, \citenamefont {Zavareh},
  \citenamefont {Salazar~Mej\'{i}a}, \citenamefont {Hofmann}, \citenamefont
  {Albert}, \citenamefont {Felser}, \citenamefont {Nicklas},\ and\
  \citenamefont {Singh}}]{Devi2018a}%
  \BibitemOpen
\bibfield  {journal} {  }\bibfield  {author} {\bibinfo {author} {\bibfnamefont
  {P.}~\bibnamefont {Devi}}, \bibinfo {author} {\bibfnamefont {M.~G.}\
  \bibnamefont {Zavareh}}, \bibinfo {author} {\bibfnamefont {C.}~\bibnamefont
  {Salazar~Mej\'{i}a}}, \bibinfo {author} {\bibfnamefont {K.}~\bibnamefont
  {Hofmann}}, \bibinfo {author} {\bibfnamefont {B.}~\bibnamefont {Albert}},
  \bibinfo {author} {\bibfnamefont {C.}~\bibnamefont {Felser}}, \bibinfo
  {author} {\bibfnamefont {M.}~\bibnamefont {Nicklas}},\ and\ \bibinfo {author}
  {\bibfnamefont {S.}~\bibnamefont {Singh}},\ }\bibfield  {title} {\bibinfo
  {title} {Reversible adiabatic temperature change in the shape memory
  {H}eusler alloy \text{Ni$_{2.2}$Mn$_{0.8}$Ga}: An effect of structural
  compatibility},\ }\href@noop {} {\bibfield  {journal} {\bibinfo  {journal}
  {Phys. Rev. Mat.}\ }\textbf {\bibinfo {volume} {2}},\ \bibinfo {pages}
  {122401(R)} (\bibinfo {year} {2018}{\natexlab{b}})}\BibitemShut {NoStop}%
\bibitem [{\citenamefont {Song}\ \emph {et~al.}(2013)\citenamefont {Song},
  \citenamefont {Chen}, \citenamefont {Dabade}, \citenamefont {Shield},\ and\
  \citenamefont {James}}]{Song2013}%
  \BibitemOpen
  \bibfield  {author} {\bibinfo {author} {\bibfnamefont {Y.}~\bibnamefont
  {Song}}, \bibinfo {author} {\bibfnamefont {X.}~\bibnamefont {Chen}}, \bibinfo
  {author} {\bibfnamefont {V.}~\bibnamefont {Dabade}}, \bibinfo {author}
  {\bibfnamefont {T.~W.}\ \bibnamefont {Shield}},\ and\ \bibinfo {author}
  {\bibfnamefont {R.~D.}\ \bibnamefont {James}},\ }\bibfield  {title} {\bibinfo
  {title} {Enhanced reversibility and unusual microstructure of a
  phase-transforming material},\ }\href@noop {} {\bibfield  {journal} {\bibinfo
   {journal} {Nature}\ }\textbf {\bibinfo {volume} {502}},\ \bibinfo {pages}
  {85} (\bibinfo {year} {2013})}\BibitemShut {NoStop}%
\bibitem [{\citenamefont {Krenke}\ \emph {et~al.}(2005)\citenamefont {Krenke},
  \citenamefont {Acet}, \citenamefont {Wasermann}, \citenamefont {Moya},
  \citenamefont {Ma\~{n}osa},\ and\ \citenamefont {Planes}}]{Krenke2005}%
  \BibitemOpen
  \bibfield  {author} {\bibinfo {author} {\bibfnamefont {T.}~\bibnamefont
  {Krenke}}, \bibinfo {author} {\bibfnamefont {M.}~\bibnamefont {Acet}},
  \bibinfo {author} {\bibfnamefont {E.~F.}\ \bibnamefont {Wasermann}}, \bibinfo
  {author} {\bibfnamefont {X.}~\bibnamefont {Moya}}, \bibinfo {author}
  {\bibfnamefont {L.}~\bibnamefont {Ma\~{n}osa}},\ and\ \bibinfo {author}
  {\bibfnamefont {A.}~\bibnamefont {Planes}},\ }\bibfield  {title} {\bibinfo
  {title} {Inverse magnetocaloric effect in ferromagnetic \text{Ni-Mn-Sn}
  alloys},\ }\href@noop {} {\bibfield  {journal} {\bibinfo  {journal} {Nat.
  Mater.}\ }\textbf {\bibinfo {volume} {4}},\ \bibinfo {pages} {450} (\bibinfo
  {year} {2005})}\BibitemShut {NoStop}%
\bibitem [{\citenamefont {Wang}\ \emph {et~al.}(2011)\citenamefont {Wang},
  \citenamefont {Liu}, \citenamefont {Ren}, \citenamefont {Xia}, \citenamefont
  {Ruan}, \citenamefont {Yi}, \citenamefont {Ding}, \citenamefont {Li},\ and\
  \citenamefont {wang}}]{Wang2011}%
  \BibitemOpen
  \bibfield  {author} {\bibinfo {author} {\bibfnamefont {B.~M.}\ \bibnamefont
  {Wang}}, \bibinfo {author} {\bibfnamefont {Y.}~\bibnamefont {Liu}}, \bibinfo
  {author} {\bibfnamefont {P.}~\bibnamefont {Ren}}, \bibinfo {author}
  {\bibfnamefont {B.}~\bibnamefont {Xia}}, \bibinfo {author} {\bibfnamefont
  {K.~B.}\ \bibnamefont {Ruan}}, \bibinfo {author} {\bibfnamefont {J.~B.}\
  \bibnamefont {Yi}}, \bibinfo {author} {\bibfnamefont {J.}~\bibnamefont
  {Ding}}, \bibinfo {author} {\bibfnamefont {X.~G.}\ \bibnamefont {Li}},\ and\
  \bibinfo {author} {\bibfnamefont {L.}~\bibnamefont {wang}},\ }\bibfield
  {title} {\bibinfo {title} {Large exchange bias after zero field cooling from
  an unmagnetized state},\ }\href@noop {} {\bibfield  {journal} {\bibinfo
  {journal} {Phys. Rev. Lett.}\ }\textbf {\bibinfo {volume} {106}},\ \bibinfo
  {pages} {077203} (\bibinfo {year} {2011})}\BibitemShut {NoStop}%
\bibitem [{\citenamefont {Nayak}\ \emph {et~al.}(2013)\citenamefont {Nayak},
  \citenamefont {Nicklas}, \citenamefont {Chadov}, \citenamefont {Shekhar},
  \citenamefont {Skourski}, \citenamefont {Winterlik},\ and\ \citenamefont
  {Felser}}]{Nayak2013}%
  \BibitemOpen
  \bibfield  {author} {\bibinfo {author} {\bibfnamefont {A.~K.}\ \bibnamefont
  {Nayak}}, \bibinfo {author} {\bibfnamefont {M.}~\bibnamefont {Nicklas}},
  \bibinfo {author} {\bibfnamefont {S.}~\bibnamefont {Chadov}}, \bibinfo
  {author} {\bibfnamefont {C.}~\bibnamefont {Shekhar}}, \bibinfo {author}
  {\bibfnamefont {Y.}~\bibnamefont {Skourski}}, \bibinfo {author}
  {\bibfnamefont {J.}~\bibnamefont {Winterlik}},\ and\ \bibinfo {author}
  {\bibfnamefont {C.}~\bibnamefont {Felser}},\ }\bibfield  {title} {\bibinfo
  {title} {Large zero-field cooled exchange-bias in bulk \text{Mn$_{2}$PtGa}},\
  }\href@noop {} {\bibfield  {journal} {\bibinfo  {journal} {Phys. Rev. Lett.}\
  }\textbf {\bibinfo {volume} {110}},\ \bibinfo {pages} {127204} (\bibinfo
  {year} {2013})}\BibitemShut {NoStop}%
\bibitem [{\citenamefont {Nayak}\ \emph {et~al.}(2015)\citenamefont {Nayak},
  \citenamefont {Nicklas}, \citenamefont {Chadov}, \citenamefont {Khuntia},
  \citenamefont {Shekhar}, \citenamefont {Kalache}, \citenamefont {Baenitz},
  \citenamefont {Skourski}, \citenamefont {Guduru}, \citenamefont {Puri},
  \citenamefont {Zeitler}, \citenamefont {Coey},\ and\ \citenamefont
  {Felser}}]{Nayak2015}%
  \BibitemOpen
  \bibfield  {author} {\bibinfo {author} {\bibfnamefont {A.~K.}\ \bibnamefont
  {Nayak}}, \bibinfo {author} {\bibfnamefont {M.}~\bibnamefont {Nicklas}},
  \bibinfo {author} {\bibfnamefont {S.}~\bibnamefont {Chadov}}, \bibinfo
  {author} {\bibfnamefont {P.}~\bibnamefont {Khuntia}}, \bibinfo {author}
  {\bibfnamefont {C.}~\bibnamefont {Shekhar}}, \bibinfo {author} {\bibfnamefont
  {A.}~\bibnamefont {Kalache}}, \bibinfo {author} {\bibfnamefont
  {M.}~\bibnamefont {Baenitz}}, \bibinfo {author} {\bibfnamefont
  {Y.}~\bibnamefont {Skourski}}, \bibinfo {author} {\bibfnamefont {V.~K.}\
  \bibnamefont {Guduru}}, \bibinfo {author} {\bibfnamefont {A.}~\bibnamefont
  {Puri}}, \bibinfo {author} {\bibfnamefont {U.}~\bibnamefont {Zeitler}},
  \bibinfo {author} {\bibfnamefont {J.~M.~D.}\ \bibnamefont {Coey}},\ and\
  \bibinfo {author} {\bibfnamefont {C.}~\bibnamefont {Felser}},\ }\bibfield
  {title} {\bibinfo {title} {Design of compensated ferrimagnetic {H}eusler
  alloys for giant tunable exchange bias},\ }\href@noop {} {\bibfield
  {journal} {\bibinfo  {journal} {Nat. Mater.}\ }\textbf {\bibinfo {volume}
  {14}},\ \bibinfo {pages} {679} (\bibinfo {year} {2015})}\BibitemShut
  {NoStop}%
\bibitem [{\citenamefont {Chatterjee}\ \emph {et~al.}(2009)\citenamefont
  {Chatterjee}, \citenamefont {Giri}, \citenamefont {De},\ and\ \citenamefont
  {Majumdar}}]{chatterjee2009}%
  \BibitemOpen
  \bibfield  {author} {\bibinfo {author} {\bibfnamefont {S.}~\bibnamefont
  {Chatterjee}}, \bibinfo {author} {\bibfnamefont {S.}~\bibnamefont {Giri}},
  \bibinfo {author} {\bibfnamefont {S.~K.}\ \bibnamefont {De}},\ and\ \bibinfo
  {author} {\bibfnamefont {S.}~\bibnamefont {Majumdar}},\ }\bibfield  {title}
  {\bibinfo {title} {Reentrant-spin-glass state
  in\text{Ni$_{2}$Mn$_{1.36}$Sn$_{0.64}$} shape-memory alloy},\ }\href@noop {}
  {\bibfield  {journal} {\bibinfo  {journal} {Phys. Rev. B}\ }\textbf {\bibinfo
  {volume} {79}},\ \bibinfo {pages} {092410} (\bibinfo {year}
  {2009})}\BibitemShut {NoStop}%
\bibitem [{\citenamefont {Singh}\ \emph {et~al.}(2012)\citenamefont {Singh},
  \citenamefont {Rawat}, \citenamefont {Muthu}, \citenamefont {D$'$Souza},
  \citenamefont {Suard}, \citenamefont {Senyshyn}, \citenamefont {Banik},
  \citenamefont {Rajput}, \citenamefont {Bhardwaj}, \citenamefont {Awasthi},
  \citenamefont {Ranjan}, \citenamefont {Arumugam}, \citenamefont {Schlagel},
  \citenamefont {Lograsso}, \citenamefont {Chakrabarti},\ and\ \citenamefont
  {Barman}}]{Singh2012}%
  \BibitemOpen
  \bibfield  {author} {\bibinfo {author} {\bibfnamefont {S.}~\bibnamefont
  {Singh}}, \bibinfo {author} {\bibfnamefont {R.}~\bibnamefont {Rawat}},
  \bibinfo {author} {\bibfnamefont {S.~E.}\ \bibnamefont {Muthu}}, \bibinfo
  {author} {\bibfnamefont {S.~W.}\ \bibnamefont {D$'$Souza}}, \bibinfo {author}
  {\bibfnamefont {E.}~\bibnamefont {Suard}}, \bibinfo {author} {\bibfnamefont
  {A.}~\bibnamefont {Senyshyn}}, \bibinfo {author} {\bibfnamefont
  {S.}~\bibnamefont {Banik}}, \bibinfo {author} {\bibfnamefont
  {P.}~\bibnamefont {Rajput}}, \bibinfo {author} {\bibfnamefont
  {S.}~\bibnamefont {Bhardwaj}}, \bibinfo {author} {\bibfnamefont {A.~M.}\
  \bibnamefont {Awasthi}}, \bibinfo {author} {\bibfnamefont {R.}~\bibnamefont
  {Ranjan}}, \bibinfo {author} {\bibfnamefont {S.}~\bibnamefont {Arumugam}},
  \bibinfo {author} {\bibfnamefont {D.~L.}\ \bibnamefont {Schlagel}}, \bibinfo
  {author} {\bibfnamefont {T.~A.}\ \bibnamefont {Lograsso}}, \bibinfo {author}
  {\bibfnamefont {A.}~\bibnamefont {Chakrabarti}},\ and\ \bibinfo {author}
  {\bibfnamefont {S.~R.}\ \bibnamefont {Barman}},\ }\bibfield  {title}
  {\bibinfo {title} {Spin valve like magnetoresistance in \text{Mn$_{2}$NiGa}
  at room temperature},\ }\href@noop {} {\bibfield  {journal} {\bibinfo
  {journal} {Phys. Rev. Lett.}\ }\textbf {\bibinfo {volume} {109}},\ \bibinfo
  {pages} {246601} (\bibinfo {year} {2012})}\BibitemShut {NoStop}%
\bibitem [{\citenamefont {Kainuma}\ \emph {et~al.}(2006)\citenamefont
  {Kainuma}, \citenamefont {Imano}, \citenamefont {Ito}, \citenamefont
  {Morito}, \citenamefont {Sutou}, \citenamefont {Oikawa}, \citenamefont
  {Fujita}, \citenamefont {Ishida}, \citenamefont {Okamoto},\ and\
  \citenamefont {Kitakami}}]{Kainuma2006}%
  \BibitemOpen
  \bibfield  {author} {\bibinfo {author} {\bibfnamefont {R.}~\bibnamefont
  {Kainuma}}, \bibinfo {author} {\bibfnamefont {Y.}~\bibnamefont {Imano}},
  \bibinfo {author} {\bibfnamefont {W.}~\bibnamefont {Ito}}, \bibinfo {author}
  {\bibfnamefont {H.}~\bibnamefont {Morito}}, \bibinfo {author} {\bibfnamefont
  {Y.}~\bibnamefont {Sutou}}, \bibinfo {author} {\bibfnamefont
  {K.}~\bibnamefont {Oikawa}}, \bibinfo {author} {\bibfnamefont
  {A.}~\bibnamefont {Fujita}}, \bibinfo {author} {\bibfnamefont
  {K.}~\bibnamefont {Ishida}}, \bibinfo {author} {\bibfnamefont
  {S.}~\bibnamefont {Okamoto}},\ and\ \bibinfo {author} {\bibfnamefont
  {K.}~\bibnamefont {Kitakami}},\ }\bibfield  {title} {\bibinfo {title}
  {Metamagnetic shape memory effect in a {H}eusler-type
  \text{Ni$_{43}$Co$_{7}$Mn$_{39}$Sn$_{11}$} polycrystalline alloy},\
  }\href@noop {} {\bibfield  {journal} {\bibinfo  {journal} {Appl. Phys.
  Lett.}\ }\textbf {\bibinfo {volume} {88}},\ \bibinfo {pages} {192513}
  (\bibinfo {year} {2006})}\BibitemShut {NoStop}%
\bibitem [{\citenamefont {Salazar~Mej\'{i}a}\ \emph {et~al.}(2015)\citenamefont
  {Salazar~Mej\'{i}a}, \citenamefont {Nayak}, \citenamefont {Schiemer},
  \citenamefont {Felser}, \citenamefont {Nicklas},\ and\ \citenamefont
  {Carpenter}}]{SalazarMejia2015}%
  \BibitemOpen
  \bibfield  {author} {\bibinfo {author} {\bibfnamefont {C.}~\bibnamefont
  {Salazar~Mej\'{i}a}}, \bibinfo {author} {\bibfnamefont {A.~K.}\ \bibnamefont
  {Nayak}}, \bibinfo {author} {\bibfnamefont {J.~A.}\ \bibnamefont {Schiemer}},
  \bibinfo {author} {\bibfnamefont {C.}~\bibnamefont {Felser}}, \bibinfo
  {author} {\bibfnamefont {M.}~\bibnamefont {Nicklas}},\ and\ \bibinfo {author}
  {\bibfnamefont {M.~A.}\ \bibnamefont {Carpenter}},\ }\bibfield  {title}
  {\bibinfo {title} {Strain behavior and lattice dynamics in
  \text{Ni$_{50}$Mn$_{35}$In$_{15}$}},\ }\href@noop {} {\bibfield  {journal}
  {\bibinfo  {journal} {J. Phys.: Condens. Matter}\ }\textbf {\bibinfo {volume}
  {27}},\ \bibinfo {pages} {415402} (\bibinfo {year} {2015})}\BibitemShut
  {NoStop}%
\bibitem [{\citenamefont {Singh}\ \emph {et~al.}(2017)\citenamefont {Singh},
  \citenamefont {Dutta}, \citenamefont {D$'$Souza}, \citenamefont {Zavareh},
  \citenamefont {Devi}, \citenamefont {Gibbs}, \citenamefont {Hickel},
  \citenamefont {Chadov}, \citenamefont {Felser},\ and\ \citenamefont
  {Pandey}}]{Singh2017}%
  \BibitemOpen
  \bibfield  {author} {\bibinfo {author} {\bibfnamefont {S.}~\bibnamefont
  {Singh}}, \bibinfo {author} {\bibfnamefont {B.}~\bibnamefont {Dutta}},
  \bibinfo {author} {\bibfnamefont {S.~W.}\ \bibnamefont {D$'$Souza}}, \bibinfo
  {author} {\bibfnamefont {M.~G.}\ \bibnamefont {Zavareh}}, \bibinfo {author}
  {\bibfnamefont {P.}~\bibnamefont {Devi}}, \bibinfo {author} {\bibfnamefont
  {A.~S.}\ \bibnamefont {Gibbs}}, \bibinfo {author} {\bibfnamefont
  {T.}~\bibnamefont {Hickel}}, \bibinfo {author} {\bibfnamefont
  {S.}~\bibnamefont {Chadov}}, \bibinfo {author} {\bibfnamefont
  {C.}~\bibnamefont {Felser}},\ and\ \bibinfo {author} {\bibfnamefont
  {D.}~\bibnamefont {Pandey}},\ }\bibfield  {title} {\bibinfo {title} {Robust
  bain distortion in the premartensite phase of a platinum-substituted
  \text{Ni$_{2}$MnGa} magnetic shape memory alloy},\ }\href@noop {} {\bibfield
  {journal} {\bibinfo  {journal} {Nat. Commun.}\ }\textbf {\bibinfo {volume}
  {8}},\ \bibinfo {pages} {1006} (\bibinfo {year} {2017})}\BibitemShut
  {NoStop}%
\bibitem [{\citenamefont {Nayak}\ \emph {et~al.}(2017)\citenamefont {Nayak},
  \citenamefont {Kumar}, \citenamefont {Ma}, \citenamefont {Werner},
  \citenamefont {Pippel}, \citenamefont {Sahoo}, \citenamefont {Damay},
  \citenamefont {R\"{o}ssler}, \citenamefont {Felser},\ and\ \citenamefont
  {Parkin}}]{Nayak2017}%
  \BibitemOpen
  \bibfield  {author} {\bibinfo {author} {\bibfnamefont {A.~K.}\ \bibnamefont
  {Nayak}}, \bibinfo {author} {\bibfnamefont {V.}~\bibnamefont {Kumar}},
  \bibinfo {author} {\bibfnamefont {T.}~\bibnamefont {Ma}}, \bibinfo {author}
  {\bibfnamefont {P.}~\bibnamefont {Werner}}, \bibinfo {author} {\bibfnamefont
  {E.}~\bibnamefont {Pippel}}, \bibinfo {author} {\bibfnamefont
  {R.}~\bibnamefont {Sahoo}}, \bibinfo {author} {\bibfnamefont
  {F.}~\bibnamefont {Damay}}, \bibinfo {author} {\bibfnamefont {U.~K.}\
  \bibnamefont {R\"{o}ssler}}, \bibinfo {author} {\bibfnamefont
  {C.}~\bibnamefont {Felser}},\ and\ \bibinfo {author} {\bibfnamefont
  {S.~S.~P.}\ \bibnamefont {Parkin}},\ }\bibfield  {title} {\bibinfo {title}
  {Magnetic antiskyrmions above room temperature in tetragonal {H}eusler
  materials},\ }\href@noop {} {\bibfield  {journal} {\bibinfo  {journal}
  {Nature}\ }\textbf {\bibinfo {volume} {548}},\ \bibinfo {pages} {561}
  (\bibinfo {year} {2017})}\BibitemShut {NoStop}%
\bibitem [{\citenamefont {Mescheriakova}\ \emph {et~al.}(2014)\citenamefont
  {Mescheriakova}, \citenamefont {Chadov}, \citenamefont {Nayak}, \citenamefont
  {R\"{o}ssler}, \citenamefont {K\"{u}bler}, \citenamefont {Andr\'{e}},
  \citenamefont {Tsirlin}, \citenamefont {Kiss}, \citenamefont {Hausdorf},
  \citenamefont {Kalache}, \citenamefont {Schnelle}, \citenamefont {Nicklas},\
  and\ \citenamefont {Felser}}]{Mescheriakova2014}%
  \BibitemOpen
  \bibfield  {author} {\bibinfo {author} {\bibfnamefont {O.}~\bibnamefont
  {Mescheriakova}}, \bibinfo {author} {\bibfnamefont {S.}~\bibnamefont
  {Chadov}}, \bibinfo {author} {\bibfnamefont {A.~K.}\ \bibnamefont {Nayak}},
  \bibinfo {author} {\bibfnamefont {U.~K.}\ \bibnamefont {R\"{o}ssler}},
  \bibinfo {author} {\bibfnamefont {J.}~\bibnamefont {K\"{u}bler}}, \bibinfo
  {author} {\bibfnamefont {G.}~\bibnamefont {Andr\'{e}}}, \bibinfo {author}
  {\bibfnamefont {A.~A.}\ \bibnamefont {Tsirlin}}, \bibinfo {author}
  {\bibfnamefont {J.}~\bibnamefont {Kiss}}, \bibinfo {author} {\bibfnamefont
  {S.}~\bibnamefont {Hausdorf}}, \bibinfo {author} {\bibfnamefont
  {A.}~\bibnamefont {Kalache}}, \bibinfo {author} {\bibfnamefont
  {W.}~\bibnamefont {Schnelle}}, \bibinfo {author} {\bibfnamefont
  {M.}~\bibnamefont {Nicklas}},\ and\ \bibinfo {author} {\bibfnamefont
  {C.}~\bibnamefont {Felser}},\ }\bibfield  {title} {\bibinfo {title} {Large
  noncollinearity and spin reorientation in the novel \text{Mn$_{2}$RhSn}
  {H}eusler magnet},\ }\href@noop {} {\bibfield  {journal} {\bibinfo  {journal}
  {Phys. Rev. Lett.}\ }\textbf {\bibinfo {volume} {113}},\ \bibinfo {pages}
  {087203} (\bibinfo {year} {2014})}\BibitemShut {NoStop}%
\bibitem [{\citenamefont {Nayak}\ \emph {et~al.}(2014)\citenamefont {Nayak},
  \citenamefont {Salazar~Mej\'{i}a}, \citenamefont {D$^{\prime}$Souza},
  \citenamefont {Chadov}, \citenamefont {Skourski}, \citenamefont {Felser},\
  and\ \citenamefont {Nicklas}}]{Nayak2014}%
  \BibitemOpen
  \bibfield  {author} {\bibinfo {author} {\bibfnamefont {A.~K.}\ \bibnamefont
  {Nayak}}, \bibinfo {author} {\bibfnamefont {C.}~\bibnamefont
  {Salazar~Mej\'{i}a}}, \bibinfo {author} {\bibfnamefont {S.~W.}\ \bibnamefont
  {D$^{\prime}$Souza}}, \bibinfo {author} {\bibfnamefont {S.}~\bibnamefont
  {Chadov}}, \bibinfo {author} {\bibfnamefont {Y.}~\bibnamefont {Skourski}},
  \bibinfo {author} {\bibfnamefont {C.}~\bibnamefont {Felser}},\ and\ \bibinfo
  {author} {\bibfnamefont {M.}~\bibnamefont {Nicklas}},\ }\bibfield  {title}
  {\bibinfo {title} {Large field-induced irreversibility in \text{Ni-Mn} based
  {H}eusler shape-memory alloys: A pulsed magnetic field study},\ }\href@noop
  {} {\bibfield  {journal} {\bibinfo  {journal} {Phys. Rev. B}\ }\textbf
  {\bibinfo {volume} {90}},\ \bibinfo {pages} {220408(R)} (\bibinfo {year}
  {2014})}\BibitemShut {NoStop}%
\bibitem [{\citenamefont {Gueltig}\ \emph {et~al.}(2014)\citenamefont
  {Gueltig}, \citenamefont {Ossmer}, \citenamefont {Ohtsuka}, \citenamefont
  {Miki}, \citenamefont {Tsuchiya}, \citenamefont {Takagi},\ and\ \citenamefont
  {Kohl}}]{Gueltig2014}%
  \BibitemOpen
  \bibfield  {author} {\bibinfo {author} {\bibfnamefont {M.}~\bibnamefont
  {Gueltig}}, \bibinfo {author} {\bibfnamefont {H.}~\bibnamefont {Ossmer}},
  \bibinfo {author} {\bibfnamefont {M.}~\bibnamefont {Ohtsuka}}, \bibinfo
  {author} {\bibfnamefont {H.}~\bibnamefont {Miki}}, \bibinfo {author}
  {\bibfnamefont {K.}~\bibnamefont {Tsuchiya}}, \bibinfo {author}
  {\bibfnamefont {T.}~\bibnamefont {Takagi}},\ and\ \bibinfo {author}
  {\bibfnamefont {M.}~\bibnamefont {Kohl}},\ }\bibfield  {title} {\bibinfo
  {title} {High frequency thermal energy harvesting using magnetic shape memory
  films},\ }\href@noop {} {\bibfield  {journal} {\bibinfo  {journal} {Adv.
  Energy Mater.}\ }\textbf {\bibinfo {volume} {4}},\ \bibinfo {pages} {1400751}
  (\bibinfo {year} {2014})}\BibitemShut {NoStop}%
\bibitem [{\citenamefont {Srivastava}\ \emph {et~al.}(2011)\citenamefont
  {Srivastava}, \citenamefont {Song}, \citenamefont {Bhatti},\ and\
  \citenamefont {James}}]{Srivastava2011}%
  \BibitemOpen
  \bibfield  {author} {\bibinfo {author} {\bibfnamefont {V.}~\bibnamefont
  {Srivastava}}, \bibinfo {author} {\bibfnamefont {Y.}~\bibnamefont {Song}},
  \bibinfo {author} {\bibfnamefont {K.}~\bibnamefont {Bhatti}},\ and\ \bibinfo
  {author} {\bibfnamefont {R.~D.}\ \bibnamefont {James}},\ }\bibfield  {title}
  {\bibinfo {title} {The direct conversion of heat to electricity using
  multiferroic alloys},\ }\href@noop {} {\bibfield  {journal} {\bibinfo
  {journal} {Adv. Energy Mater.}\ }\textbf {\bibinfo {volume} {1}},\ \bibinfo
  {pages} {97} (\bibinfo {year} {2011})}\BibitemShut {NoStop}%
\bibitem [{\citenamefont {Singh}\ \emph {et~al.}(2015)\citenamefont {Singh},
  \citenamefont {Kushwaha}, \citenamefont {Scheibel}, \citenamefont {Liermann},
  \citenamefont {Barman}, \citenamefont {Acet}, \citenamefont {Felser},\ and\
  \citenamefont {Pandey}}]{Singh2015}%
  \BibitemOpen
  \bibfield  {author} {\bibinfo {author} {\bibfnamefont {S.}~\bibnamefont
  {Singh}}, \bibinfo {author} {\bibfnamefont {P.}~\bibnamefont {Kushwaha}},
  \bibinfo {author} {\bibfnamefont {F.}~\bibnamefont {Scheibel}}, \bibinfo
  {author} {\bibfnamefont {H.~P.}\ \bibnamefont {Liermann}}, \bibinfo {author}
  {\bibfnamefont {S.~R.}\ \bibnamefont {Barman}}, \bibinfo {author}
  {\bibfnamefont {M.}~\bibnamefont {Acet}}, \bibinfo {author} {\bibfnamefont
  {C.}~\bibnamefont {Felser}},\ and\ \bibinfo {author} {\bibfnamefont
  {D.}~\bibnamefont {Pandey}},\ }\bibfield  {title} {\bibinfo {title} {Residual
  stress induced stabilization of martensite phase and its effect on the
  magetostructural transition in \text{Mn}-rich \text{Ni-Mn-In/Ga} magnetic
  shape memory alloys},\ }\href@noop {} {\bibfield  {journal} {\bibinfo
  {journal} {Phys. Rev. B}\ }\textbf {\bibinfo {volume} {92}},\ \bibinfo
  {pages} {02015} (\bibinfo {year} {2015})}\BibitemShut {NoStop}%
\bibitem [{\citenamefont {Gottschall}\ \emph {et~al.}(2015)\citenamefont
  {Gottschall}, \citenamefont {Skokov}, \citenamefont {Frincu},\ and\
  \citenamefont {Gutfleisch}}]{Gottschall2015}%
  \BibitemOpen
  \bibfield  {author} {\bibinfo {author} {\bibfnamefont {T.}~\bibnamefont
  {Gottschall}}, \bibinfo {author} {\bibfnamefont {K.~P.}\ \bibnamefont
  {Skokov}}, \bibinfo {author} {\bibfnamefont {B.}~\bibnamefont {Frincu}},\
  and\ \bibinfo {author} {\bibfnamefont {O.}~\bibnamefont {Gutfleisch}},\
  }\bibfield  {title} {\bibinfo {title} {Large reversible magnetocaloric effect
  in \text{Ni-Mn-In-Co}},\ }\href@noop {} {\bibfield  {journal} {\bibinfo
  {journal} {Appl. Phys. Lett.}\ }\textbf {\bibinfo {volume} {106}},\ \bibinfo
  {pages} {021901} (\bibinfo {year} {2015})}\BibitemShut {NoStop}%
\bibitem [{\citenamefont {Khovaylo}\ \emph {et~al.}(2009)\citenamefont
  {Khovaylo}, \citenamefont {Kanomata}, \citenamefont {Tanaka}, \citenamefont
  {Nakashima}, \citenamefont {Amako}, \citenamefont {Kainuma}, \citenamefont
  {Umetsu}, \citenamefont {Morito},\ and\ \citenamefont {Miki}}]{Khovaylo2009}%
  \BibitemOpen
  \bibfield  {author} {\bibinfo {author} {\bibfnamefont {V.~V.}\ \bibnamefont
  {Khovaylo}}, \bibinfo {author} {\bibfnamefont {T.}~\bibnamefont {Kanomata}},
  \bibinfo {author} {\bibfnamefont {T.}~\bibnamefont {Tanaka}}, \bibinfo
  {author} {\bibfnamefont {M.}~\bibnamefont {Nakashima}}, \bibinfo {author}
  {\bibfnamefont {Y.}~\bibnamefont {Amako}}, \bibinfo {author} {\bibfnamefont
  {R.}~\bibnamefont {Kainuma}}, \bibinfo {author} {\bibfnamefont {R.~Y.}\
  \bibnamefont {Umetsu}}, \bibinfo {author} {\bibfnamefont {H.}~\bibnamefont
  {Morito}},\ and\ \bibinfo {author} {\bibfnamefont {H.}~\bibnamefont {Miki}},\
  }\bibfield  {title} {\bibinfo {title} {Magnetic properties of
  \text{Ni$_{50}$Mn$_{34.8}$In$_{15.2}$ probed by M\"{o}ssbauer
  spectroscopy}},\ }\href@noop {} {\bibfield  {journal} {\bibinfo  {journal}
  {Phys. Rev. B}\ }\textbf {\bibinfo {volume} {80}},\ \bibinfo {pages} {144409}
  (\bibinfo {year} {2009})}\BibitemShut {NoStop}%
\bibitem [{\citenamefont {Oikawa}\ \emph {et~al.}(2006)\citenamefont {Oikawa},
  \citenamefont {Ito}, \citenamefont {Imano}, \citenamefont {Kainuma},
  \citenamefont {Ishida}, \citenamefont {Okamoto}, \citenamefont {Kitakami},\
  and\ \citenamefont {Kanomata}}]{Oikawa2006}%
  \BibitemOpen
  \bibfield  {author} {\bibinfo {author} {\bibfnamefont {K.}~\bibnamefont
  {Oikawa}}, \bibinfo {author} {\bibfnamefont {W.}~\bibnamefont {Ito}},
  \bibinfo {author} {\bibfnamefont {Y.}~\bibnamefont {Imano}}, \bibinfo
  {author} {\bibfnamefont {R.}~\bibnamefont {Kainuma}}, \bibinfo {author}
  {\bibfnamefont {K.}~\bibnamefont {Ishida}}, \bibinfo {author} {\bibfnamefont
  {S.}~\bibnamefont {Okamoto}}, \bibinfo {author} {\bibfnamefont
  {O.}~\bibnamefont {Kitakami}},\ and\ \bibinfo {author} {\bibfnamefont
  {T.}~\bibnamefont {Kanomata}},\ }\bibfield  {title} {\bibinfo {title} {Effect
  of magnetic field on martensitic transition of
  \text{Ni$_{46}$Mn$_{41}$In$_{13}$} {H}eusler alloy},\ }\href@noop {}
  {\bibfield  {journal} {\bibinfo  {journal} {Appl. Phys. Lett.}\ }\textbf
  {\bibinfo {volume} {88}},\ \bibinfo {pages} {122507} (\bibinfo {year}
  {2006})}\BibitemShut {NoStop}%
\bibitem [{\citenamefont {Sharma}\ \emph {et~al.}(2010)\citenamefont {Sharma},
  \citenamefont {Chattopadhyay}, \citenamefont {Khandelwal},\ and\
  \citenamefont {Roy}}]{Sharma2010}%
  \BibitemOpen
  \bibfield  {author} {\bibinfo {author} {\bibfnamefont {V.~K.}\ \bibnamefont
  {Sharma}}, \bibinfo {author} {\bibfnamefont {M.~K.}\ \bibnamefont
  {Chattopadhyay}}, \bibinfo {author} {\bibfnamefont {A.}~\bibnamefont
  {Khandelwal}},\ and\ \bibinfo {author} {\bibfnamefont {S.~B.}\ \bibnamefont
  {Roy}},\ }\bibfield  {title} {\bibinfo {title} {Martensitic transition near
  room temperature and the temperature- and magnetic-field-induced
  multifunctional properties of \text{Ni$_{49}$CuMn$_{34}$In$_{16}$} alloy},\
  }\href@noop {} {\bibfield  {journal} {\bibinfo  {journal} {Phys. Rev. B}\
  }\textbf {\bibinfo {volume} {82}},\ \bibinfo {pages} {172411} (\bibinfo
  {year} {2010})}\BibitemShut {NoStop}%
\bibitem [{\citenamefont {Ye}\ \emph {et~al.}(2010)\citenamefont {Ye},
  \citenamefont {Kimura}, \citenamefont {Miura}, \citenamefont {Shirai},
  \citenamefont {Cui}, \citenamefont {Shimada}, \citenamefont {Namatame},
  \citenamefont {Taniguchi}, \citenamefont {Ueda}, \citenamefont {Kobayashi},
  \citenamefont {Kainuma}, \citenamefont {Shishido}, \citenamefont
  {Fukushima},\ and\ \citenamefont {Kanomata}}]{Ye2010}%
  \BibitemOpen
  \bibfield  {author} {\bibinfo {author} {\bibfnamefont {M.}~\bibnamefont
  {Ye}}, \bibinfo {author} {\bibfnamefont {A.}~\bibnamefont {Kimura}}, \bibinfo
  {author} {\bibfnamefont {Y.}~\bibnamefont {Miura}}, \bibinfo {author}
  {\bibfnamefont {M.}~\bibnamefont {Shirai}}, \bibinfo {author} {\bibfnamefont
  {Y.~T.}\ \bibnamefont {Cui}}, \bibinfo {author} {\bibfnamefont
  {K.}~\bibnamefont {Shimada}}, \bibinfo {author} {\bibfnamefont
  {H.}~\bibnamefont {Namatame}}, \bibinfo {author} {\bibfnamefont
  {M.}~\bibnamefont {Taniguchi}}, \bibinfo {author} {\bibfnamefont
  {S.}~\bibnamefont {Ueda}}, \bibinfo {author} {\bibfnamefont {K.}~\bibnamefont
  {Kobayashi}}, \bibinfo {author} {\bibfnamefont {R.}~\bibnamefont {Kainuma}},
  \bibinfo {author} {\bibfnamefont {T.}~\bibnamefont {Shishido}}, \bibinfo
  {author} {\bibfnamefont {K.}~\bibnamefont {Fukushima}},\ and\ \bibinfo
  {author} {\bibfnamefont {T.}~\bibnamefont {Kanomata}},\ }\bibfield  {title}
  {\bibinfo {title} {Role of electronic structure in the martensitic phase
  transition of \text{Ni$_{2}$Mn$_{1+x}$Sn$_{1-x}$} studied by hard-x-ray
  photoelectron spectroscopy and ab-initio calculation},\ }\href@noop {}
  {\bibfield  {journal} {\bibinfo  {journal} {Phys. Rev. Lett}\ }\textbf
  {\bibinfo {volume} {104}},\ \bibinfo {pages} {176401} (\bibinfo {year}
  {2010})}\BibitemShut {NoStop}%
\bibitem [{\citenamefont {Khan}\ \emph {et~al.}(2016)\citenamefont {Khan},
  \citenamefont {Brock},\ and\ \citenamefont {Sugerman}}]{Khan2016}%
  \BibitemOpen
  \bibfield  {author} {\bibinfo {author} {\bibfnamefont {M.}~\bibnamefont
  {Khan}}, \bibinfo {author} {\bibfnamefont {J.}~\bibnamefont {Brock}},\ and\
  \bibinfo {author} {\bibfnamefont {I.}~\bibnamefont {Sugerman}},\ }\bibfield
  {title} {\bibinfo {title} {Anomalous transport properties of
  \text{Ni$_{2}$Mn$_{1-x}$Cr$_{x}$Ga} heusler alloys at the
  martensite-austenite phase transition},\ }\href@noop {} {\bibfield  {journal}
  {\bibinfo  {journal} {Phys. Rev. B}\ }\textbf {\bibinfo {volume} {93}},\
  \bibinfo {pages} {054419} (\bibinfo {year} {2016})}\BibitemShut {NoStop}%
\bibitem [{\citenamefont {Ma\~{n}osa}\ \emph {et~al.}(2008)\citenamefont
  {Ma\~{n}osa}, \citenamefont {Moya}, \citenamefont {Planes}, \citenamefont
  {Gutfleisch}, \citenamefont {Lyubina}, \citenamefont {Barrio}, \citenamefont
  {Tamarit}, \citenamefont {Aksoy}, \citenamefont {Krenke},\ and\ \citenamefont
  {Acet}}]{Manosa2008}%
  \BibitemOpen
  \bibfield  {author} {\bibinfo {author} {\bibfnamefont {L.}~\bibnamefont
  {Ma\~{n}osa}}, \bibinfo {author} {\bibfnamefont {X.}~\bibnamefont {Moya}},
  \bibinfo {author} {\bibfnamefont {A.}~\bibnamefont {Planes}}, \bibinfo
  {author} {\bibfnamefont {O.}~\bibnamefont {Gutfleisch}}, \bibinfo {author}
  {\bibfnamefont {J.}~\bibnamefont {Lyubina}}, \bibinfo {author} {\bibfnamefont
  {M.}~\bibnamefont {Barrio}}, \bibinfo {author} {\bibfnamefont {J.~L.}\
  \bibnamefont {Tamarit}}, \bibinfo {author} {\bibfnamefont {S.}~\bibnamefont
  {Aksoy}}, \bibinfo {author} {\bibfnamefont {T.}~\bibnamefont {Krenke}},\ and\
  \bibinfo {author} {\bibfnamefont {M.}~\bibnamefont {Acet}},\ }\bibfield
  {title} {\bibinfo {title} {Effect of hydrostatic pressure on the magnetism
  and martensitic transition of \text{Ni-Mn-In} magnetic superelastic alloys},\
  }\href@noop {} {\bibfield  {journal} {\bibinfo  {journal} {Appl. Phys.
  Lett.}\ }\textbf {\bibinfo {volume} {92}},\ \bibinfo {pages} {012515}
  (\bibinfo {year} {2008})}\BibitemShut {NoStop}%
\bibitem [{\citenamefont {Eiling}\ and\ \citenamefont
  {Schilling}(1981)}]{Eiling1981}%
  \BibitemOpen
  \bibfield  {author} {\bibinfo {author} {\bibfnamefont {A.}~\bibnamefont
  {Eiling}}\ and\ \bibinfo {author} {\bibfnamefont {J.~S.}\ \bibnamefont
  {Schilling}},\ }\bibfield  {title} {\bibinfo {title} {Pressure and
  temperature dependence of electrical resistivity of {P}b and {S}n from
  \text{$1-300$ K} and 0-10 {GP}a-use as continuous resistive pressure monitor
  accurate over wide temperature range; superconductivity under pressure in
  {P}b, {S}n and {I}n},\ }\href@noop {} {\bibfield  {journal} {\bibinfo
  {journal} {J. Phys. F: Met. Phys.}\ }\textbf {\bibinfo {volume} {11}},\
  \bibinfo {pages} {623} (\bibinfo {year} {1981})}\BibitemShut {NoStop}%
\bibitem [{\citenamefont {Wollmann}\ \emph {et~al.}(2017)\citenamefont
  {Wollmann}, \citenamefont {Nayak}, \citenamefont {Parkin},\ and\
  \citenamefont {Felser}}]{Wollmann2017}%
  \BibitemOpen
  \bibfield  {author} {\bibinfo {author} {\bibfnamefont {L.}~\bibnamefont
  {Wollmann}}, \bibinfo {author} {\bibfnamefont {A.~K.}\ \bibnamefont {Nayak}},
  \bibinfo {author} {\bibfnamefont {S.~S.~P.}\ \bibnamefont {Parkin}},\ and\
  \bibinfo {author} {\bibfnamefont {C.}~\bibnamefont {Felser}},\ }\bibfield
  {title} {\bibinfo {title} {Heusler 4.0: Tunable materials},\ }\href@noop {}
  {\bibfield  {journal} {\bibinfo  {journal} {Ann. Rev. Mater. Res.}\ }\textbf
  {\bibinfo {volume} {47}},\ \bibinfo {pages} {247} (\bibinfo {year}
  {2017})}\BibitemShut {NoStop}%
\bibitem [{\citenamefont {Li}\ \emph {et~al.}(2019)\citenamefont {Li},
  \citenamefont {Yang}, \citenamefont {Li}, \citenamefont {Li}, \citenamefont
  {Yang}, \citenamefont {Yan}, \citenamefont {Vald\'{e}s}, \citenamefont
  {Llamazares}, \citenamefont {Zhang}, \citenamefont {Esling},\ and\
  \citenamefont {Zuo}}]{Li2019}%
  \BibitemOpen
  \bibfield  {author} {\bibinfo {author} {\bibfnamefont {Z.}~\bibnamefont
  {Li}}, \bibinfo {author} {\bibfnamefont {J.}~\bibnamefont {Yang}}, \bibinfo
  {author} {\bibfnamefont {D.}~\bibnamefont {Li}}, \bibinfo {author}
  {\bibfnamefont {Z.}~\bibnamefont {Li}}, \bibinfo {author} {\bibfnamefont
  {B.}~\bibnamefont {Yang}}, \bibinfo {author} {\bibfnamefont {H.}~\bibnamefont
  {Yan}}, \bibinfo {author} {\bibfnamefont {C.~F.~S.}\ \bibnamefont
  {Vald\'{e}s}}, \bibinfo {author} {\bibfnamefont {J.~L.~S.}\ \bibnamefont
  {Llamazares}}, \bibinfo {author} {\bibfnamefont {Y.}~\bibnamefont {Zhang}},
  \bibinfo {author} {\bibfnamefont {C.}~\bibnamefont {Esling}},\ and\ \bibinfo
  {author} {\bibfnamefont {L.}~\bibnamefont {Zuo}},\ }\bibfield  {title}
  {\bibinfo {title} {Tuning the reversible magnetocaloric effect in
  \text{Ni-Mn-In}-based alloys through co and cu co-doping},\ }\href@noop {}
  {\bibfield  {journal} {\bibinfo  {journal} {Adv. Elect. Mater.}\ }\textbf
  {\bibinfo {volume} {5}},\ \bibinfo {pages} {1800845} (\bibinfo {year}
  {2019})}\BibitemShut {NoStop}%
\bibitem [{\citenamefont {Qu}\ \emph {et~al.}(2019)\citenamefont {Qu},
  \citenamefont {Condal}, \citenamefont {Ma\~{n}osa}, \citenamefont {Planes},
  \citenamefont {Cong}, \citenamefont {Nie}, \citenamefont {Ren},\ and\
  \citenamefont {Wang}}]{Qu2019}%
  \BibitemOpen
  \bibfield  {author} {\bibinfo {author} {\bibfnamefont {Y.}~\bibnamefont
  {Qu}}, \bibinfo {author} {\bibfnamefont {A.~G.}\ \bibnamefont {Condal}},
  \bibinfo {author} {\bibfnamefont {L.}~\bibnamefont {Ma\~{n}osa}}, \bibinfo
  {author} {\bibfnamefont {A.}~\bibnamefont {Planes}}, \bibinfo {author}
  {\bibfnamefont {D.}~\bibnamefont {Cong}}, \bibinfo {author} {\bibfnamefont
  {Z.}~\bibnamefont {Nie}}, \bibinfo {author} {\bibfnamefont {Y.}~\bibnamefont
  {Ren}},\ and\ \bibinfo {author} {\bibfnamefont {Y.}~\bibnamefont {Wang}},\
  }\bibfield  {title} {\bibinfo {title} {Outstanding caloric performances for
  energy-efficient multicaloric cooling in a \text{Ni-Mn}-based multifunctional
  alloy},\ }\href@noop {} {\bibfield  {journal} {\bibinfo  {journal} {Acta
  Mater.}\ }\textbf {\bibinfo {volume} {177}},\ \bibinfo {pages} {46} (\bibinfo
  {year} {2019})}\BibitemShut {NoStop}%
\bibitem [{\citenamefont {Liu}\ \emph {et~al.}(2019)\citenamefont {Liu},
  \citenamefont {You}, \citenamefont {Huang}, \citenamefont {Batashev},
  \citenamefont {Maschek}, \citenamefont {Gong}, \citenamefont {Miao},
  \citenamefont {Xu}, \citenamefont {van Dijk},\ and\ \citenamefont
  {Br\"uck}}]{Liu19}%
  \BibitemOpen
  \bibfield  {author} {\bibinfo {author} {\bibfnamefont {J.}~\bibnamefont
  {Liu}}, \bibinfo {author} {\bibfnamefont {X.}~\bibnamefont {You}}, \bibinfo
  {author} {\bibfnamefont {B.}~\bibnamefont {Huang}}, \bibinfo {author}
  {\bibfnamefont {I.}~\bibnamefont {Batashev}}, \bibinfo {author}
  {\bibfnamefont {M.}~\bibnamefont {Maschek}}, \bibinfo {author} {\bibfnamefont
  {Y.}~\bibnamefont {Gong}}, \bibinfo {author} {\bibfnamefont {X.}~\bibnamefont
  {Miao}}, \bibinfo {author} {\bibfnamefont {F.}~\bibnamefont {Xu}}, \bibinfo
  {author} {\bibfnamefont {N.}~\bibnamefont {van Dijk}},\ and\ \bibinfo
  {author} {\bibfnamefont {E.}~\bibnamefont {Br\"uck}},\ }\bibfield  {title}
  {\bibinfo {title} {Reversible low-field magnetocaloric effect in
  {Ni-Mn-In}-based {Heusler} alloys},\ }\href@noop {} {\bibfield  {journal}
  {\bibinfo  {journal} {Phys. Rev. Mat.}\ }\textbf {\bibinfo {volume} {3}},\
  \bibinfo {pages} {084409} (\bibinfo {year} {2019})}\BibitemShut {NoStop}%
\bibitem [{\citenamefont {Stern-Taulats}\ \emph {et~al.}(2014)\citenamefont
  {Stern-Taulats}, \citenamefont {Castillo-Villa}, \citenamefont {Ma\~{n}osa},
  \citenamefont {Frontera}, \citenamefont {Pramanick}, \citenamefont
  {Majumdar},\ and\ \citenamefont {Planes}}]{Stern14}%
  \BibitemOpen
  \bibfield  {author} {\bibinfo {author} {\bibfnamefont {E.}~\bibnamefont
  {Stern-Taulats}}, \bibinfo {author} {\bibfnamefont {P.~O.}\ \bibnamefont
  {Castillo-Villa}}, \bibinfo {author} {\bibfnamefont {L.}~\bibnamefont
  {Ma\~{n}osa}}, \bibinfo {author} {\bibfnamefont {C.}~\bibnamefont
  {Frontera}}, \bibinfo {author} {\bibfnamefont {S.}~\bibnamefont {Pramanick}},
  \bibinfo {author} {\bibfnamefont {S.}~\bibnamefont {Majumdar}},\ and\
  \bibinfo {author} {\bibfnamefont {A.}~\bibnamefont {Planes}},\ }\bibfield
  {title} {\bibinfo {title} {Magnetocaloric effect in the low hysteresis
  {Ni-Mn-In} metamagnetic shape-memory {Heusler} alloy},\ }\href@noop {}
  {\bibfield  {journal} {\bibinfo  {journal} {J. Appl. Phys.}\ }\textbf
  {\bibinfo {volume} {115}},\ \bibinfo {pages} {173907} (\bibinfo {year}
  {2014})}\BibitemShut {NoStop}%
\bibitem [{\citenamefont {Gottschall}\ \emph {et~al.}(2016)\citenamefont
  {Gottschall}, \citenamefont {Skokov}, \citenamefont {Benke}, \citenamefont
  {Gruner}, \citenamefont {Ghorbani~Zavareh},\ and\ \citenamefont
  {Gutfleisch}}]{Gottschall2016a}%
  \BibitemOpen
  \bibfield  {author} {\bibinfo {author} {\bibfnamefont {T.}~\bibnamefont
  {Gottschall}}, \bibinfo {author} {\bibfnamefont {K.~P.}\ \bibnamefont
  {Skokov}}, \bibinfo {author} {\bibfnamefont {D.}~\bibnamefont {Benke}},
  \bibinfo {author} {\bibfnamefont {M.~E.}\ \bibnamefont {Gruner}}, \bibinfo
  {author} {\bibfnamefont {M.}~\bibnamefont {Ghorbani~Zavareh}},\ and\ \bibinfo
  {author} {\bibfnamefont {O.}~\bibnamefont {Gutfleisch}},\ }\bibfield  {title}
  {\bibinfo {title} {Contradictory role of the magnetic contribution in inverse
  magnetocaloric {H}eusler materials},\ }\href@noop {} {\bibfield  {journal}
  {\bibinfo  {journal} {Phys. Rev. B.}\ }\textbf {\bibinfo {volume} {93}},\
  \bibinfo {pages} {184431} (\bibinfo {year} {2016})}\BibitemShut {NoStop}%
\bibitem [{\citenamefont {Salazar~Mej\'{i}a}\ \emph {et~al.}(2017)\citenamefont
  {Salazar~Mej\'{i}a}, \citenamefont {K\"{u}chler}, \citenamefont {Nayak},
  \citenamefont {Skourski}, \citenamefont {Wosnitza}, \citenamefont {Felser},\
  and\ \citenamefont {Nicklas}}]{Salazar2017}%
  \BibitemOpen
  \bibfield  {author} {\bibinfo {author} {\bibfnamefont {C.}~\bibnamefont
  {Salazar~Mej\'{i}a}}, \bibinfo {author} {\bibfnamefont {R.}~\bibnamefont
  {K\"{u}chler}}, \bibinfo {author} {\bibfnamefont {A.~K.}\ \bibnamefont
  {Nayak}}, \bibinfo {author} {\bibfnamefont {Y.}~\bibnamefont {Skourski}},
  \bibinfo {author} {\bibfnamefont {J.}~\bibnamefont {Wosnitza}}, \bibinfo
  {author} {\bibfnamefont {C.}~\bibnamefont {Felser}},\ and\ \bibinfo {author}
  {\bibfnamefont {M.}~\bibnamefont {Nicklas}},\ }\bibfield  {title} {\bibinfo
  {title} {Uniaxial-stress tuned large magnetic-shape-memory effect in
  \text{Ni-Co-Mn-Sb} {H}eusler alloys},\ }\href@noop {} {\bibfield  {journal}
  {\bibinfo  {journal} {Appl. Phys. Lett.}\ }\textbf {\bibinfo {volume}
  {110}},\ \bibinfo {pages} {071901} (\bibinfo {year} {2017})}\BibitemShut
  {NoStop}%
\bibitem [{\citenamefont {Li}\ \emph {et~al.}(2010)\citenamefont {Li},
  \citenamefont {Jing}, \citenamefont {Zhang}, \citenamefont {Yu},
  \citenamefont {Chen}, \citenamefont {Kang}, \citenamefont {Cao},\ and\
  \citenamefont {Zhang}}]{Li2010}%
  \BibitemOpen
  \bibfield  {author} {\bibinfo {author} {\bibfnamefont {Z.}~\bibnamefont
  {Li}}, \bibinfo {author} {\bibfnamefont {C.}~\bibnamefont {Jing}}, \bibinfo
  {author} {\bibfnamefont {H.~L.}\ \bibnamefont {Zhang}}, \bibinfo {author}
  {\bibfnamefont {D.~H.}\ \bibnamefont {Yu}}, \bibinfo {author} {\bibfnamefont
  {L.}~\bibnamefont {Chen}}, \bibinfo {author} {\bibfnamefont {B.~J.}\
  \bibnamefont {Kang}}, \bibinfo {author} {\bibfnamefont {S.~X.}\ \bibnamefont
  {Cao}},\ and\ \bibinfo {author} {\bibfnamefont {J.~C.}\ \bibnamefont
  {Zhang}},\ }\bibfield  {title} {\bibinfo {title} {A large and reproducible
  metamagnetic shape memory effect in polycrystalline
  \text{Ni$_{45}$Co$_{5}$Mn$_{37}$In$_{13}$} {H}eusler alloy},\ }\href@noop {}
  {\bibfield  {journal} {\bibinfo  {journal} {J. Appl. Phys.}\ }\textbf
  {\bibinfo {volume} {108}},\ \bibinfo {pages} {113908} (\bibinfo {year}
  {2010})}\BibitemShut {NoStop}%
\bibitem [{\citenamefont {Ito}\ \emph {et~al.}(2007)\citenamefont {Ito},
  \citenamefont {Imano}, \citenamefont {Kainuma}, \citenamefont {Oikawa},\ and\
  \citenamefont {Ishida}}]{Ito2007}%
  \BibitemOpen
  \bibfield  {author} {\bibinfo {author} {\bibfnamefont {W.}~\bibnamefont
  {Ito}}, \bibinfo {author} {\bibfnamefont {Y.}~\bibnamefont {Imano}}, \bibinfo
  {author} {\bibfnamefont {R.}~\bibnamefont {Kainuma}}, \bibinfo {author}
  {\bibfnamefont {K.}~\bibnamefont {Oikawa}},\ and\ \bibinfo {author}
  {\bibfnamefont {K.}~\bibnamefont {Ishida}},\ }\bibfield  {title} {\bibinfo
  {title} {Martensitic and magnetic transformation behaviors in {H}eusler-type
  \text{NiMnIn} and \text{NiCoMnIn} metamagnetic shape memory alloys},\
  }\href@noop {} {\bibfield  {journal} {\bibinfo  {journal} {Metall. and Mater.
  Trans. A}\ }\textbf {\bibinfo {volume} {38A}},\ \bibinfo {pages} {759}
  (\bibinfo {year} {2007})}\BibitemShut {NoStop}%
\bibitem [{\citenamefont {Gottschall}\ \emph {et~al.}(2017)\citenamefont
  {Gottschall}, \citenamefont {Taulats}, \citenamefont {Ma\~{n}osa},
  \citenamefont {Planes}, \citenamefont {Skokov},\ and\ \citenamefont
  {Gutfleisch}}]{Gottschall17}%
  \BibitemOpen
  \bibfield  {author} {\bibinfo {author} {\bibfnamefont {T.}~\bibnamefont
  {Gottschall}}, \bibinfo {author} {\bibfnamefont {E.~S.}\ \bibnamefont
  {Taulats}}, \bibinfo {author} {\bibfnamefont {L.}~\bibnamefont {Ma\~{n}osa}},
  \bibinfo {author} {\bibfnamefont {A.}~\bibnamefont {Planes}}, \bibinfo
  {author} {\bibfnamefont {K.~P.}\ \bibnamefont {Skokov}},\ and\ \bibinfo
  {author} {\bibfnamefont {O.}~\bibnamefont {Gutfleisch}},\ }\bibfield  {title}
  {\bibinfo {title} {Reversibility of minor hysteresis loops in magnetocaloric
  {H}eusler alloys},\ }\href@noop {} {\bibfield  {journal} {\bibinfo  {journal}
  {Appl. Phys. Lett.}\ }\textbf {\bibinfo {volume} {110}},\ \bibinfo {pages}
  {223904} (\bibinfo {year} {2017})}\BibitemShut {NoStop}%
\bibitem [{\citenamefont {Engelbrecht}\ \emph {et~al.}(2013)\citenamefont
  {Engelbrecht}, \citenamefont {Nielsen}, \citenamefont {Bahl}, \citenamefont
  {Carroll},\ and\ \citenamefont {Asten}}]{Engelbrecht13}%
  \BibitemOpen
  \bibfield  {author} {\bibinfo {author} {\bibfnamefont {K.}~\bibnamefont
  {Engelbrecht}}, \bibinfo {author} {\bibfnamefont {K.~K.}\ \bibnamefont
  {Nielsen}}, \bibinfo {author} {\bibfnamefont {C.~R.~H.}\ \bibnamefont
  {Bahl}}, \bibinfo {author} {\bibfnamefont {C.~P.}\ \bibnamefont {Carroll}},\
  and\ \bibinfo {author} {\bibfnamefont {D.~V.}\ \bibnamefont {Asten}},\
  }\bibfield  {title} {\bibinfo {title} {Material properties and modeling
  characteristics for \text{MnFeP$_{1−x}$As$_{x}$} materials for application
  in magnetic refrigeration},\ }\href@noop {} {\bibfield  {journal} {\bibinfo
  {journal} {J. Appl. Phys.}\ }\textbf {\bibinfo {volume} {113}},\ \bibinfo
  {pages} {173510} (\bibinfo {year} {2013})}\BibitemShut {NoStop}%
\bibitem [{\citenamefont {Guillou}\ \emph {et~al.}(2014)\citenamefont
  {Guillou}, \citenamefont {Yibole}, \citenamefont {Porcari}, \citenamefont
  {Zhang}, \citenamefont {Dijk},\ and\ \citenamefont {Br\"{u}ck}}]{Guillou14}%
  \BibitemOpen
  \bibfield  {author} {\bibinfo {author} {\bibfnamefont {F.}~\bibnamefont
  {Guillou}}, \bibinfo {author} {\bibfnamefont {H.}~\bibnamefont {Yibole}},
  \bibinfo {author} {\bibfnamefont {G.}~\bibnamefont {Porcari}}, \bibinfo
  {author} {\bibfnamefont {L.}~\bibnamefont {Zhang}}, \bibinfo {author}
  {\bibfnamefont {N.~H.~V.}\ \bibnamefont {Dijk}},\ and\ \bibinfo {author}
  {\bibfnamefont {E.}~\bibnamefont {Br\"{u}ck}},\ }\bibfield  {title} {\bibinfo
  {title} {Magnetocaloric effect, cyclability and coefficient of refrigerant
  performance in the \text{MnFe(P, Si, B)} system},\ }\href@noop {} {\bibfield
  {journal} {\bibinfo  {journal} {J. Appl. Phys.}\ }\textbf {\bibinfo {volume}
  {116}},\ \bibinfo {pages} {063903} (\bibinfo {year} {2014})}\BibitemShut
  {NoStop}%
\bibitem [{\citenamefont {Skokov}\ \emph {et~al.}(2013)\citenamefont {Skokov},
  \citenamefont {Moore}, \citenamefont {Liu}, \citenamefont {Karpenkov},
  \citenamefont {Krautz},\ and\ \citenamefont {Gutfleisch}}]{Skokov13}%
  \BibitemOpen
  \bibfield  {author} {\bibinfo {author} {\bibfnamefont {K.~P.}\ \bibnamefont
  {Skokov}}, \bibinfo {author} {\bibfnamefont {J.~D.}\ \bibnamefont {Moore}},
  \bibinfo {author} {\bibfnamefont {J.}~\bibnamefont {Liu}}, \bibinfo {author}
  {\bibfnamefont {A.~Y.}\ \bibnamefont {Karpenkov}}, \bibinfo {author}
  {\bibfnamefont {M.}~\bibnamefont {Krautz}},\ and\ \bibinfo {author}
  {\bibfnamefont {O.}~\bibnamefont {Gutfleisch}},\ }\bibfield  {title}
  {\bibinfo {title} {Influence of thermal hysteresis and field cycling on the
  magnetocaloric effect in \text{LaFe$_{11.6}$Si$_{1.4}$}},\ }\href@noop {}
  {\bibfield  {journal} {\bibinfo  {journal} {J. Alloys Compd.}\ }\textbf
  {\bibinfo {volume} {552}},\ \bibinfo {pages} {310} (\bibinfo {year}
  {2013})}\BibitemShut {NoStop}%
\end{thebibliography}%

\end{document}